\documentclass[
  prb,
  aps,
  twocolumn,
  nolongbibliography,
  superscriptaddress,
  showpacs
]{revtex4-2}

\usepackage[%
  colorlinks=true,
  linkcolor=black, 
  citecolor=blue, 
  urlcolor=blue, 
  unicode=true
]{hyperref}

\usepackage{graphicx,amssymb,amsmath,braket,lipsum,xcolor}
\usepackage{mathtools,mathrsfs}

\begin{document}


\title{Metastable antiphase boundary ordering in CaFe$_{2}$O$_{4}$}

\author{H. Lane}
\affiliation{School of Physics and Astronomy, University of Edinburgh, Edinburgh EH9 3JZ, United Kingdom}
\affiliation{School of Chemistry and Centre for Science at Extreme Conditions, University of Edinburgh, Edinburgh EH9 3FJ, United Kingdom}
\affiliation{ISIS Pulsed Neutron and Muon Source, STFC Rutherford Appleton Laboratory, Harwell Campus, Didcot, Oxon, OX11 0QX, United Kingdom}
\author{E. E. Rodriguez}
\affiliation{Department of Chemistry and Biochemistry, University of Maryland, College Park, Maryland 20742, USA}
\author{H. C. Walker}
\affiliation{ISIS Pulsed Neutron and Muon Source, STFC Rutherford Appleton Laboratory, Harwell Campus, Didcot, Oxon, OX11 0QX, United Kingdom}
\author{Ch. Niedermayer}
\affiliation{Laboratory for Neutron Scattering, Paul Scherrer Institut, CH-5232 Villigen, Switzerland}
\author{U. Stuhr}
\affiliation{Laboratory for Neutron Scattering, Paul Scherrer Institut, CH-5232 Villigen, Switzerland}
\author{R. I. Bewley}
\affiliation{ISIS Pulsed Neutron and Muon Source, STFC Rutherford Appleton Laboratory, Harwell Campus, Didcot, Oxon, OX11 0QX, United Kingdom}
\author{D. J. Voneshen}
\affiliation{ISIS Pulsed Neutron and Muon Source, STFC Rutherford Appleton Laboratory, Harwell Campus, Didcot, Oxon, OX11 0QX, United Kingdom}
\author{M. A. Green}
\affiliation{School of Physical Sciences, University of Kent, Canterbury CT2 7NH, United Kingdom}
\author{J. A. Rodriguez-Rivera}
\affiliation{NIST Center for Neutron Research, National Institute of Standards and Technology, Gaithersburg, MD, USA}
\affiliation{Department of Materials Science and Engineering, University of Maryland, College Park, MD, USA}
\author{P. Fouquet}
\affiliation{Institute Laue-Langevin, 6 rue Jules Horowitz, Boite Postale 156, 38042 Grenoble Cedex 9, France}
\author{S.-W. Cheong}
\affiliation{Rutgers Center for Emergent Materials and Department of Physics and Astronomy, Rutgers University, Piscataway, New Jersey 08854, USA}
\author{J. P. Attfield}
\affiliation{School of Chemistry and Centre for Science at Extreme Conditions, University of Edinburgh, Edinburgh EH9 3FJ, United Kingdom}
\author{R. A. Ewings}
\affiliation{ISIS Pulsed Neutron and Muon Source, STFC Rutherford Appleton Laboratory, Harwell Campus, Didcot, Oxon, OX11 0QX, United Kingdom}
\author{C. Stock}
\affiliation{School of Physics and Astronomy, University of Edinburgh, Edinburgh EH9 3JZ, United Kingdom}
\date{\today}

\begin{abstract}
CaFe$_{2}$O$_{4}$ is an $S=5/2$ antiferromagnet exhibiting two magnetic orders which shows regions of coexistence at some temperatures. Using a Green's function formalism, we model neutron scattering data of the spin wave excitations in this material, ellucidating the microscopic spin Hamiltonian. In doing so, we suggest that the low temperature A phase order $(\uparrow\uparrow\downarrow\downarrow)$ finds its origins in the freezing of antiphase boundaries created by thermal fluctuations in a parent B phase order $(\uparrow\downarrow\uparrow\downarrow)$. The low temperature magnetic order observed in CaFe$_{2}$O$_{4}$ is thus the result of a competition between the exchange coupling along $c$, which favors the B phase, and the single-ion anisotropy which stabilizes thermally-generated antiphase boundaries, leading to static metastable A phase order at low temperatures. 
\end{abstract}


\maketitle

\section{Introduction}
The manipulation of the domain-wall motion of ferromagnets via a coupling to external fields has been suggested as a promising mechanism for the design of logic gates \cite{Allwood} and racetrack memory devices \cite{Parkin,Ho} for the next generation of quantum devices. Additional attention has been paid to the control of antiferromagnetic domain walls, which overcome the practical difficulties of the large stray fields associated with their ferromagnetic counterparts, yet cannot be controlled with a simple external field \cite{Baltz}. Nonetheless, mechanisms have been suggested for the control of antiferromagnetic domain walls ranging from thermal activation \cite{Kim2} to spin-orbital torques \cite{Zelezny} and magnon-driving \cite{Kim1}. 

One antiferromagnetic system which may prove instructive in the study of magnon-soliton interactions and antiphase boundary effects is the $S=\frac{5}{2}$ antiferromagnet CaFe$_{2}$O$_{4}$. Polarized neutron diffraction data show the existence of spatially extended Bloch walls separating antiphase regions of antiferromagnetic order \cite{Orphan}. The antiphase boundaries have been found to carry an uncompensated local moment and are hence tunable in field \cite{Orphan}. Furthermore, the low energy magnetic excitation spectrum was found to exhibit discrete modes \cite{Solitary}, perhaps indicative of confinement of solitons within a nonlinear potential which arises due to frustration between domains on weakly coupled chains \cite{Lane}. Knowledge of the microscopic spin Hamiltonian is a necessity before a full understanding of the antiphase boundaries can be gained, yet a consistent picture of the magnetic interactions in CaFe$_{2}$O$_{4}$ has proved elusive.        

CaFe$_{2}$O$_{4}$ exhibits two magnetically ordered phases. The high temperature B phase, consists of two-dimensional networks of coupled zig-zag chains which are stacked along $c$ in the $(\uparrow \downarrow \uparrow \downarrow)$ pattern (see Fig. \ref{structure fig} $(a)$). As temperature is decreased, the A phase develops, which differs only in its $(\uparrow \uparrow \downarrow \downarrow)$ $c$ axis stacking \cite{Corliss,Solitary}. These two phases are observed to coexist, yet the temperature range of this coexistence and the ultimate low temperature structure differ between single crystal and powder samples \cite{Das,Songvilay}. In single crystals the ordered moment does not saturate at $5\mu_{B}$, with spectral weight redistributed to momentum-broadened rods of scattering along $c^{*}$, confirmed by polarization analysis to be magnetic in origin \cite{Solitary}, indicative of antiphase boundaries along $c$. From polarized diffraction data of the momentum-broadened component, at T = 5 K the correlation length along $c$ was determined to be $\sim 1-2$ unit cells \cite{Solitary} indicating highly localized correlations. The ability to measure magnetic diffuse scattering in powders without polarization analysis or a large amount of diffuse spectral weight is limited, making the presence of antiphase boundaries in powder samples difficult to detect. However, the ordered moment is observed to be suppressed in polycrystalline samples \cite{Songvilay}. A full characterization of the magnetic excitations at both high and low temperature has not yet been presented. 

In this paper, we address the nature of the magnetic order in CaFe$_{2}$O$_{4}$ and offer an explanation for the differing behavior observed in powders and single crystals. The low temperature A phase is shown to be metastable, with short range correlations, analogous to the field-induced metastable states recently reported in CoV$_{2}$O$_{6}$ \cite{Edwards}, in which antiphase boundaries order to form a new phase with a different translational symmetry \cite{Fisher}. These arguments are supported by neutron scattering data at both high and low temperatures, demonstrating the nature of the magnetic fluctuations in single crystal CaFe$_{2}$O$_{4}$. Finally, using a random phase approximation (RPA) Green's function formalism, we model the magnetic excitations in CaFe$_{2}$O$_{4}$ and determine the spin Hamiltonian.

\section{Antiphase Boundaries}
\label{APB}
\subsection{Structure}
\label{structure}
\begin{figure*}
\includegraphics[width=1\linewidth]{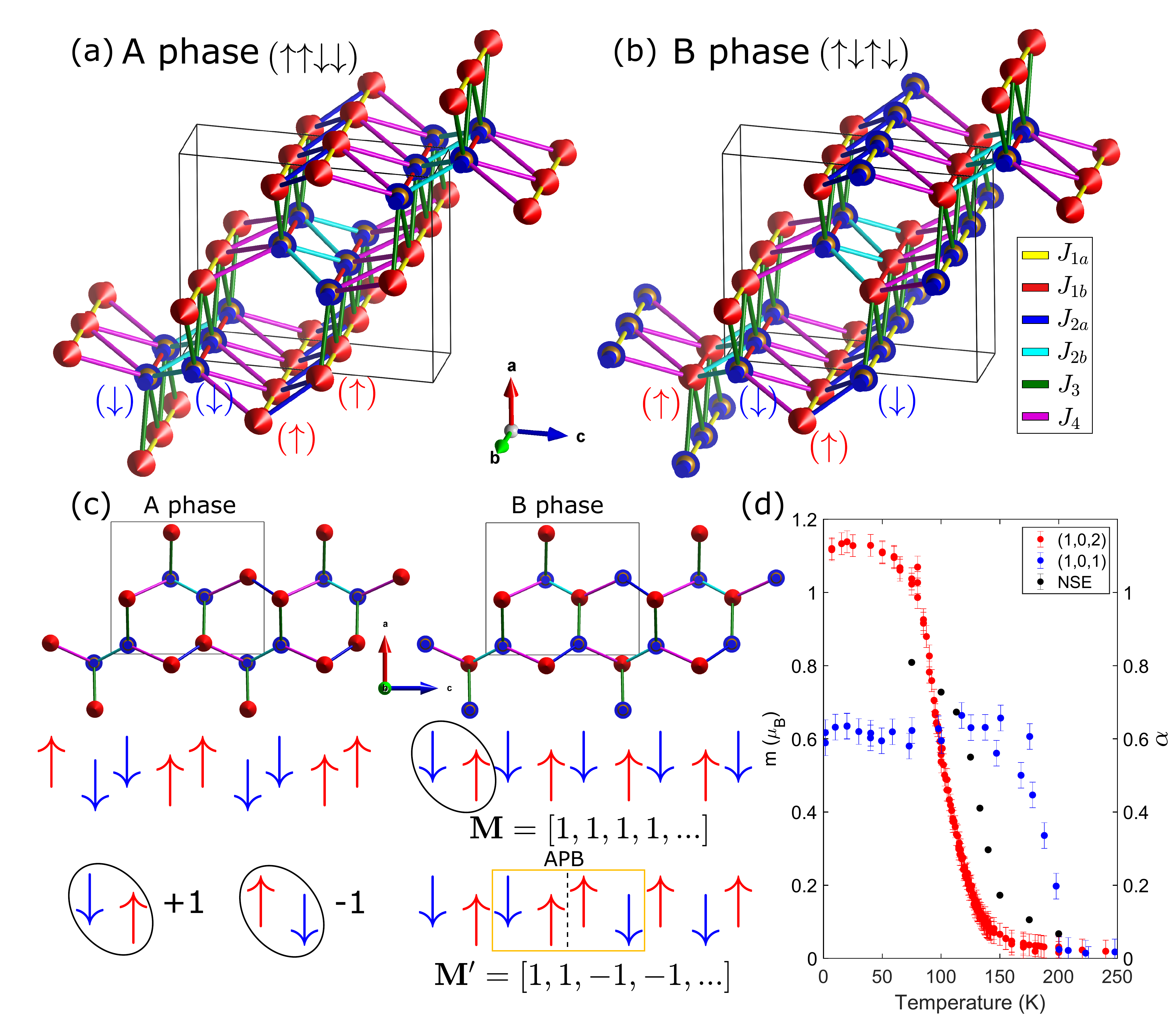}
\caption{\label{structure fig} Structure of CaFe$_{2}$O$_{4}$ in the A $(a)$ and B $(b)$ phases, which differ in their $c$ axis stacking. The couplings $J_{1a}$ and $J_{1b}$ link parallel spins along $b$, with the $J_{2a}$ and $J_{2b}$ linking parallel and antiparallel spins in the A and B phases respectively. The exchanges $J_{3}$ and $J_{4}$ define antiferromagnetically aligned chains in both phases. $(c)$ Magnetic structure along the $c$ axis showing the $(\uparrow\uparrow\downarrow\downarrow)$ and $(\uparrow\downarrow\uparrow\downarrow)$ configuration of the A and B phases respectively. The effect of an antiphase boundary (APB) in the B phase order is illustrated, giving rise to local A phase order (golden rectangle). The mapping onto the matrix $\mathbf{M}$ is demonstrated, with each spin pair mapped onto $\pm1$. $(d)$ Temperature dependence of the order parameters of the A and B phases, Q=(1,0,2) and Q=(1,0,1) respectively. Black points show the temperature dependence of the static component, $\alpha$,  of neutron spin echo (NSE) at Q=(1,0,1.5), data reproduced from Ref. \cite{Solitary}. }
\end{figure*}

CaFe$_{2}$O$_{4}$ crystallizes in the orthorhombic $Pnma$ space group ($a$=9.230\AA, $b$=3.017\AA, $c$=10.689\AA), with coupled zig-zag chains of Fe$^{3+}$ (S=5/2, L=0) ions in the $a-b$ plane \cite{Decker,Hill,Corliss,Allain}. Previous studies have reported the stabilization of two competing magnetic orders below $T_{N} \approx$ 200 K \cite{Bertaut,Corliss,Solitary,Das,Songvilay}. In the low temperature A phase (Fig. \ref{structure fig} $(a)$) the stacking along $c$ is $(\uparrow\uparrow\downarrow\downarrow)$ with the couplings $J_{2a}$ and $J_{2b}$ connecting parallel spins. In the high temperature B phase (Fig. \ref{structure fig} $(b)$), the $c$ axis stacking is $(\uparrow\downarrow\uparrow\downarrow)$ with $J_{2a}$ and $J_{2b}$ coupling antiparallel spins.    

\begin{figure*}
\includegraphics[width=1\linewidth]{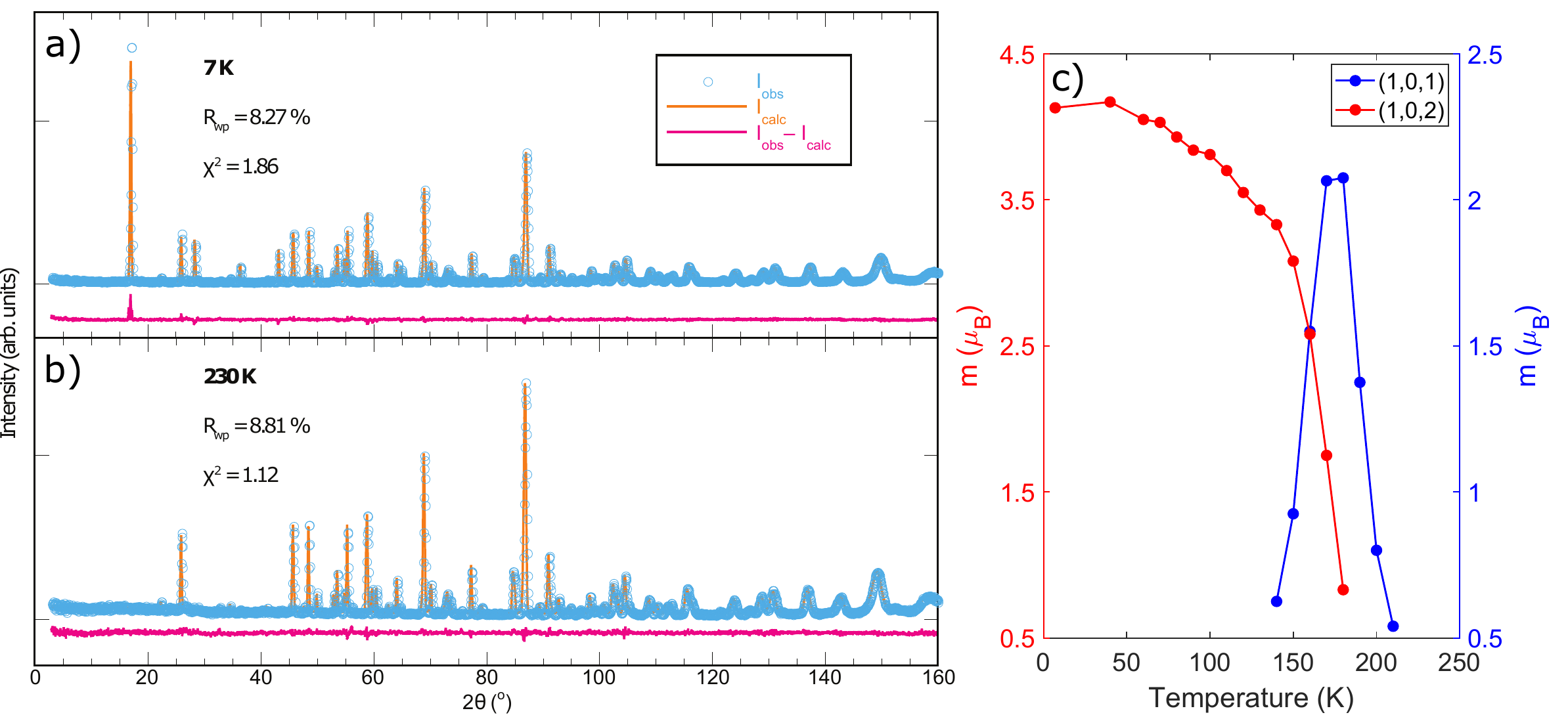}
\caption{\label{powder} $(a)$ Neutron diffraction data of a powder sample of CaFe$_{2}$O$_{4}$ measured on BT-1 at $(a)$ T = 7 K and $(b)$ T = 230 K. $(c)$ Magnetic moment of the $Q=(1,0,1)$ and $Q=(1,0,2)$ Bragg peaks in the powder sample. The small window of coexistence and loss of the B phase at low temperature shows a qualitatively different behavior to the single crystal sample \cite{Solitary}.}
\end{figure*}

The behavior observed is qualitatively different between powder and single crystal samples and is summarized in Table \ref{Magnetic structures}. To examine the magnetic structure of the powder samples, we used the BT-1 diffractometer at the NIST Center for Neutron Research (NCNR) with wavelength $\lambda$=2.0782 \AA (Ge 311 monochromator). The low and high temperature diffraction patterns are shown in Fig. \ref{powder} $(a)$ and $(b)$ respectively. In the powder samples, B phase order, as indicated by the presence of the $Q=(1,0,1)$ peak, is observed on cooling below T $\approx$ 200 K and is maximal at T $\approx$ 175 K. Below this temperature, the $Q=(1,0,2)$ peak begins to accumulate spectral weight, overtaking the B phase in intensity at T $\approx$ 150 K (Fig. \ref{powder} $(c)$). The B phase peaks are observed to disappear at around T = 125 K after a brief temperature window of coexistence between T $\approx$ 125 K and T $\approx$ 175 K. The powder diffraction data would thus indicate a preferential A phase ordering at low temperature, with $(\uparrow\uparrow\downarrow\downarrow)$ stacking along $c$, rather than the B phase order with its $(\uparrow\downarrow\uparrow\downarrow)$ arrangement. This result is in agreement with the findings of Songvilay \textit{et al} \cite{Songvilay} who have further shown that chemical doping with Cr prevents the stabilization of A phase order, observing only the B phase in CaCr$_{0.5}$Fe$_{1.5}$O$_{4}$. Pure CaCr$_{2}$O$_{4}$ shows an altogether different magnetic structure, with an incommensurate cycloidal propagation vector \cite{CaCr1,CaCr2,CaCr3}.    

The story is somewhat different in single-crystal samples. Between T $\approx$ 200 K and T $\approx$ 100  K, B phase magnetic order is dominant as seen in Fig. \ref{structure fig} $(d)$. This is confirmed by magnetic susceptibility measurements \cite{Das} showing a feature at the onset of B phase order. Below T $\approx$ 100K the A phase becomes the more prevalent magnetic order (Fig. \ref{structure fig} $(d)$). The existence of a transition to the A phase is argued on the grounds of the appearance of the peak at $Q=(1,0,2)$, however it is important to note that no thermodynamic measurements have been reported showing the existence of a second phase transition, and only a single order parameter was detected in the M{\"o}ssbauer spectroscopy measurements, showing a power law temperature dependence \cite{Damerio}. Observing the relative intensities of the  $(1,0,2)$ and $(1,0,1)$ peaks, the two phases can be seen to saturate in a 2:1 ratio at low temperature (Fig. \ref{structure fig} $(d)$). 

It has previously been suggested that the phase coexistence could originate from a fine balancing of the exchange parameter $J_{2a/b}$ on the ferromagnetic/antiferromagnetic threshold which is sensitive to subtle changes in the bond angle as a function of temperature \cite{Songvilay}. However, these arguments rely on an element of exchange disorder to account for the persistence of B phase order down to low temperatures in the single crystal samples, and the region of phase coexistence in the high temperature phase of the powder. Moreover, the discrepancy between the powder and single crystal data remains unexplained. We now present an alternative explanation for the temperature dependence of the phase coexistence based on antiphase domain formation.   

\begin{table}[h]
\caption{\label{Magnetic structures} Magnetic structures observed in CaFe$_{2}$O$_{4}$. Data from Cr doped sample reproduced from Ref. \cite{Songvilay}. }
\begin{ruledtabular}
\begin{tabular}{cccc}
Sample &B phase & A phase & Coexistence? \\ 
\hline
CaFe$_{2}$O$_{4}$ single crystal & 200-5 K & 175-5 K  & 175-5 K \\ 
CaFe$_{2}$O$_{4}$ powder & 200-150 K & 175-5 K & 175-150 K  \\ 
CaCr$_{0.5}$Fe$_{1.5}$O$_{4}$ powder & 200-5 K & $\times$ & $\times$  \\ 

\end{tabular}
\end{ruledtabular}
\end{table}   

We first discuss the single crystal before turning our attention to the powder samples. The single crystals described in this paper are the same as those used in Refs. \cite{Solitary,Orphan,Lane} and were grown using a mirror furnace, as described in the Supplemental Material of Ref. \cite{Solitary}. Previous studies of these single crystals \cite{Orphan,Solitary} have demonstrated the presence of rods of diffuse magnetic scattering along [0,0,L] indicating that correlations along $c$ are short-range. Furthermore, at T = 200 K neutron spin echo (NSE) measurements reveal the dynamical nature of this diffuse scattering \cite{Solitary,Orphan}, with the static component increasing rapidly as the sample was cooled below T = 100 K (Fig. \ref{structure fig} $(d)$). By examining the magnetic structure in the two phases, we can see that the creation of an antiphase domain boundary in global B phase order gives rise to a local A phase stacking (and vice versa) as demonstrated in Fig. \ref{structure fig} $(c)$ \cite{Orphan}. Consequently, we argue that the phase coexistence at low temperature can be understood as arising due to the freezing-in of antiphase boundaries in a parent B phase order. In order to demonstrate this, we introduce the following toy model of domain formation. 

The chain along $c$ is split into pairs of spins, with each pair assigned the value of +1 or -1 depending on the orientation of the spin pair (Fig. \ref{structure fig} $(c)$). In a pure B phase arrangement the magnetic structure can be represented by the infinite array, $\mathbf{M}=\pm[1,1,1,1,...]$. Where $\pm$ labels the two degenerate spin configurations (of which we select the positive state from now on, for definiteness). To introduce a domain wall at position $i$, we flip the signs from position $i+1$ onwards, $\mathbf{M}=[1,1,-1,-1,...]$, for example. Let $\mathcal{F}(i,p)$ be the operator that has a $p$(\%) chance of flipping the signs on all sites after site $i$, hence creating an antiphase boundary at site $i$. After operating on each element of the array  with the nonlocal operator 
\begin{equation}
\mathbf{M}'=\prod_{i}^{N}\mathcal{F}(i,p)\mathbf{M},
\end{equation}
\noindent we then analyze the local order by examining the relative signs on each site. Each occurrence of the pattern $\mathbf{M}'=[...,\pm 1,\pm 1,...]$ can be assigned to the B phase, with $\mathbf{M}'=[...,\pm 1,\mp 1,...]$ belonging to the A phase, leading to an array of length $N-1$, $\mathbf{P}=[A,B,B,A,...]$, for example, describing the local order. 

For $p=67$\%, the ratios of A to B phase are found to be 2:1 in agreement with the low temperature neutron diffraction data \cite{Solitary}. The B phase magnetic unit cell consists of along two spins along $c$ (or one element of $\mathbf{M}$) and hence a flipping ratio of $p=67$\% gives rise to domains with an average size of 1.5 elements, corresponding to 1.5 unit cells, in agreement with the measured correlation length of 1-2 unit cells, from neutron diffraction \cite{Solitary}. The same analysis can be applied to a parent A phase order, with $p=33\%$, leading to an average domain size of four elements of $\mathbf{M}$ which, again corresponds to a correlation length of 1.5 unit cells (owing to the doubled magnetic unit cell of the A phase). However, the appearance of the $Q=(1,0,1)$ peak at $T_{N}$ along with a significant component of magnetic dynamical diffuse scattering, which becomes static on the onset of A phase order is suggestive of the former scenario. We thus conclude that the single crystal data are consistent with a parent B phase order with antiphase domain boundaries that freeze-in at low temperature leading to local A phase order. This is still suggestive of a small $J_{2a/b}$ so that the energy cost of creating an antiphase boundary is of the order of the temperature, but we conclude that this bond must be strictly antiferromagnetic, in order that the parent magnetic order is B phase. 

The persistence of the B phase $(1,0,1)$ peak down to low temperature in the single crystal sample is indicative of domain pinning effects arising due the presence of oxygen vacancies, known to be present in CaFe$_{2}$O$_{4}$ single crystals \footnotetext[1]{To our knowledge, no samples with excess oxygen have been reported.}\cite{Das,Note1}. If the flipping ratio were to tend towards $p=100\%$, we would expect pure A phase order at low temperature and a disappearance of the B phase Bragg peaks, precisely as observed in the CaFe$_{2}$O$_{4}$ powder samples \cite{Songvilay}. This is to be expected if the powder samples were to have fewer oxygen vacancies and hence facilitate the full conversion of B phase to A phase order. The magnetization measurements of Das \textit{et al} \cite{Das} demonstate that vacancy-driven disorder alone cannot account for the discrepancy between the powder and single crystal samples, indeed another ingredient is needed. 

Crucial to the survival of the low temperature A phase is the presence of an anisotropy gap that stabilizes the A phase structure at low temperature despite the frustration of $J_{2a/b}$. We note that the neutron scattering measurements of Songvilay \textit{et al} suggest a significant reduction in the anisotropy gap in the Cr doped samples \cite{Songvilay}, which may explain the failure to stabilize A phase order at low temperatures. In the $3d^{5}$ high spin complexes, owing to the absence of an orbital moment, the spin Hamiltonian is expected to be isotropic. The observed anisotropy gap is thus evidence of the mixing of higher energy multiplets into the ground state orbital singlet, $^{6}S$. This mixing occurs due to higher order processes such as a second order process involving the spin-spin interaction and an axial crystal field via the $^{6}D$ state \cite{Pryce} or to fourth order via squares of the spin-orbit and distortion terms \cite{Watanabe}. Ultimately, the anisotropy terms that appear in the effective spin Hamiltonian must respect the crystal symmetry \cite{Bleaney} and hence should be proportional to the Steven's parameter, $\mu\sim B_{2}^{0}$ \cite{Yosida,AbragamBleaney}. The vital role of the axial distortion term in mixing higher order multiplets into the $^{6}S$ ground state indicates that the anisotropy should couple strongly to strain in CaFe$_{2}$O$_{4}$, which exhibits a significant distortion of the local octahedral environment \cite{Decker}. It is therefore unsurprising that doping suppresses the magnitude of the anisotropy \cite{Songvilay} and that the magnetic behavior shows a strong dependence on the density of oxygen vacancies \cite{Das}, since both processes affect the local axial crystal field. The role that anisotropy plays in the stabilization of the A phase in turn suggests an explanation for the differing behavior in powder and single crystal samples. In powder samples, the grinding process introduces strain which would indicate an enhancement of the single ion anisotropy parameter, $\mu$, promoting the stabilization of A phase order. We now further explore the temperature dependence of the anisotropy gap in single crystal samples using neutron scattering. 

\subsection{Anisotropy}
\label{Tdep}
\begin{figure}
\includegraphics[width=1\linewidth]{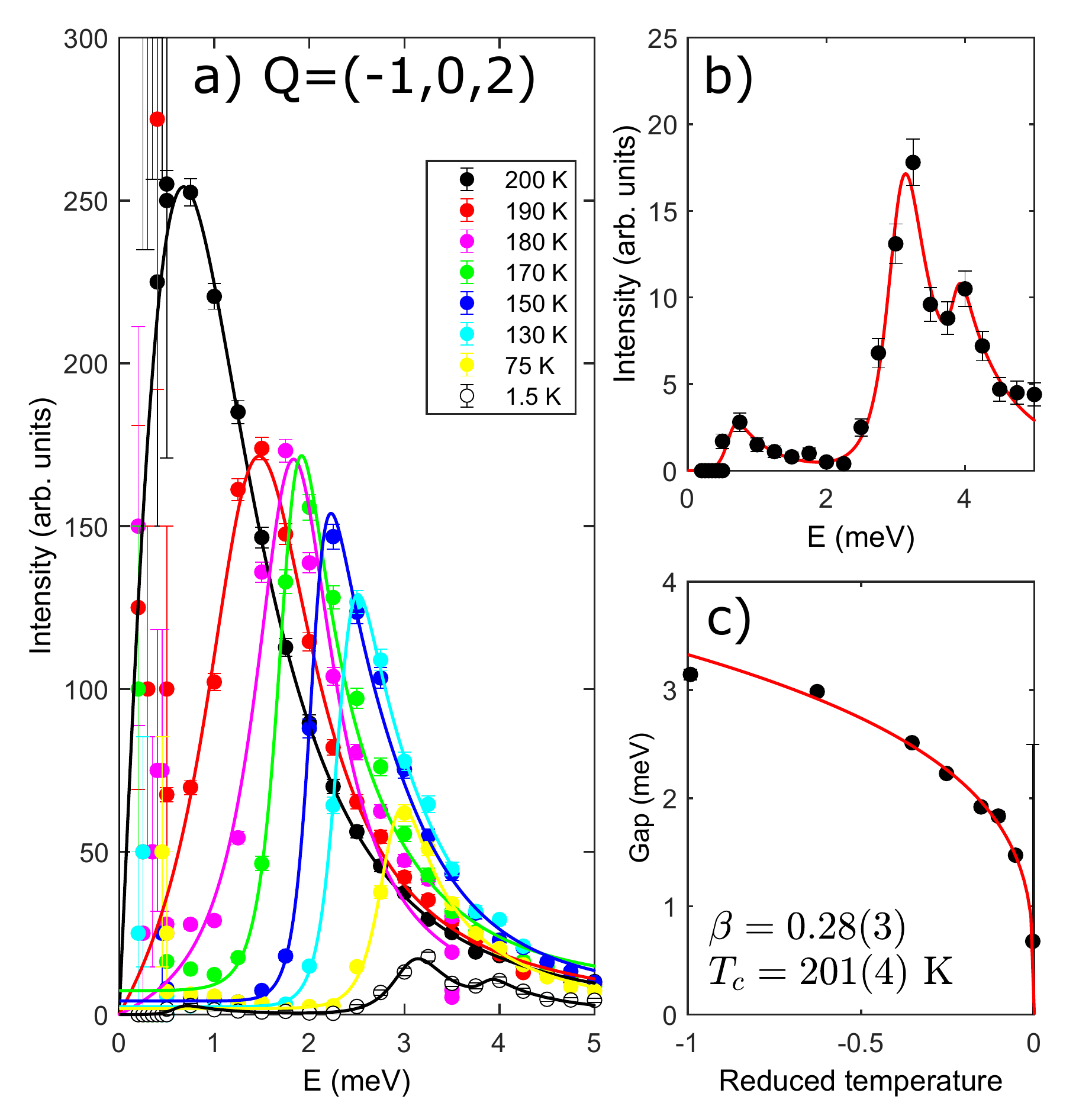}
\caption{\label{gap}$(a)$ Temperature dependence of the gap at $Q=(-1,0,2)$ measured on the RITA-II spectrometer. Solid lines are fits to asymmetric Lorentzians. $(b)$ Anisotropy gap at T=1.5 K with $\Delta=3.14(5)$meV. A low intensity peak at 1 meV is seen originating from in-gap mode, discussed in Ref. \cite{Orphan} in addition to a discrete mode at 3.9 meV originating from non-classical excitations \cite{Solitary,Orphan,Lane}. $(c)$ Extracted magnitudes of the gap as a function of reduced temperature, $t$. The data have been corrected for the Bose-Einstein population factor at each temperature.}
\end{figure}

Temperature-dependent constant Q scans at the $Q=(-1,0,2)$ position were conducted on the RITA-II triple-axis spectrometer at the Paul Scherrer Institute (Villigen, Switzerland) \cite{Lefmann} (Fig. \ref{gap} $(a)$). The asymmetric lineshape arises from the finite resolution and the curvature of the dispersion curve. The peaks are resolution-limited and for convenience we approximate the asymmetric lineshape with an antisymmetric Lorentzian function,

\begin{equation}\label{eq:flux_dissp}
I (E) \propto [n(\omega)+1] \left({1 \over {1+\left({{E-\Omega_{0}} \over \Gamma}\right)^{2}}}- {1 \over {1+\left({{E+\Omega_{0}} \over \Gamma}\right)^{2}}}\right) 
\end{equation}

\noindent whose peak width is allowed to vary sigmoidally 

\begin{equation}
\Gamma(E)=\frac{2\Gamma_{0}}{1+\textrm{exp}\left[a(E-\Omega_{0})\right]}
\end{equation}

\noindent such that the degree of asymmetry is controlled by a single parameter, $a$ and for $a=0$, the width becomes symmetric, $\Gamma=\Gamma_{0}$ \cite{Stancik}. The value of the asymmetry parameter, along with $\Gamma_{0}$, $\Omega_{0}$ and an overall scaling factor were fitted using the \texttt{HORACE} package \cite{Horace}.  The value of the gap follows a power law behavior, vanishing above T $\approx$ 200 K, concomitant with the loss of order along $c$. This is in good agreement with the temperature at which the $Q=(1,0,1)$ peak vanishes (Fig. \ref{structure fig} $(d)$). At T = 1.5 K a second peak is seen above the gap $\sim$ 4 meV, in Fig. \ref{gap} $(b)$, which can be understood in terms of the discrete non-classical excitations reported previously in this system \cite{Solitary,Orphan,Lane}. By plotting the gap as a function of reduced temperature $t=(T-T_{c})/T_{c}$, we can fit a dimensionless scaling exponent and critical temperature according to $\Delta \sim |t|^{\beta}$. The fitted value of $\beta=0.28(3)$ is below the expected scaling exponent for the 3d Ising model ($\beta=0.3265(15)$) \cite{Pelissetto}. The departure of the gap's scaling exponent from the expected critical exponent of the order parameter indicates the presence of some temperature dependence of the anisotropy parameter, beyond a simple renormalization due to a thermal fluctuation-driven reduction of the magnetization, and hence a decoupling of the magnetic order parameter and the anisotropy parameter. Such a temperature dependence has been observed in other ferrites and materials exhibiting strong magnetostrictive effects \cite{Shenker,Bozorth,Garcia}. The ramifications of this temperature dependence of the anisotropy parameter will be discussed further later in the paper. We now analyze the phonon excitations which would be sensitive to any structural domains.  

\subsection{Acoustic phonons}
The spin-wave analysis presented here shows that the magnetic excitations in CaFe$_{2}$O$_{4}$ can be consistently understood in terms of the same exchange constants in the high (B phase) and low (A phase) temperature phases, up to a small temperature renormalization.  To confirm the lack of temperature dependent structural effects, we discuss the transverse acoustic phonons.

The lifetime and energy positions of acoustic phonons are sensitive to the formation of structural defects or localized structural domains.   This has been shown in scattering studies of, for example,  localized polar domains~\cite{Burns83:48} in relaxor ferroelectrics such as Pb(Zn,Mg)$_{1/3}$Nb$_{2/3}$O$_{3}$~\cite{Koo02:65,Stock04:69,Stock05:74,Stock12:86} and the disordered perovskite K$_{1-x}$Li$_{x}$TaO$_{3}$~\cite{Stock14:90}.   To confirm the lack of any structural domains forming that may drive either the antiphase boundaries discussed above or the transition from the B phase to the A phase, on cooling, we investigated the temperature dependence of transverse acoustic phonons propagating both along $c$ and $a$ axes in CaFe$_{2}$O$_{4}$.
\begin{figure}[b]
  \includegraphics[width=95mm]{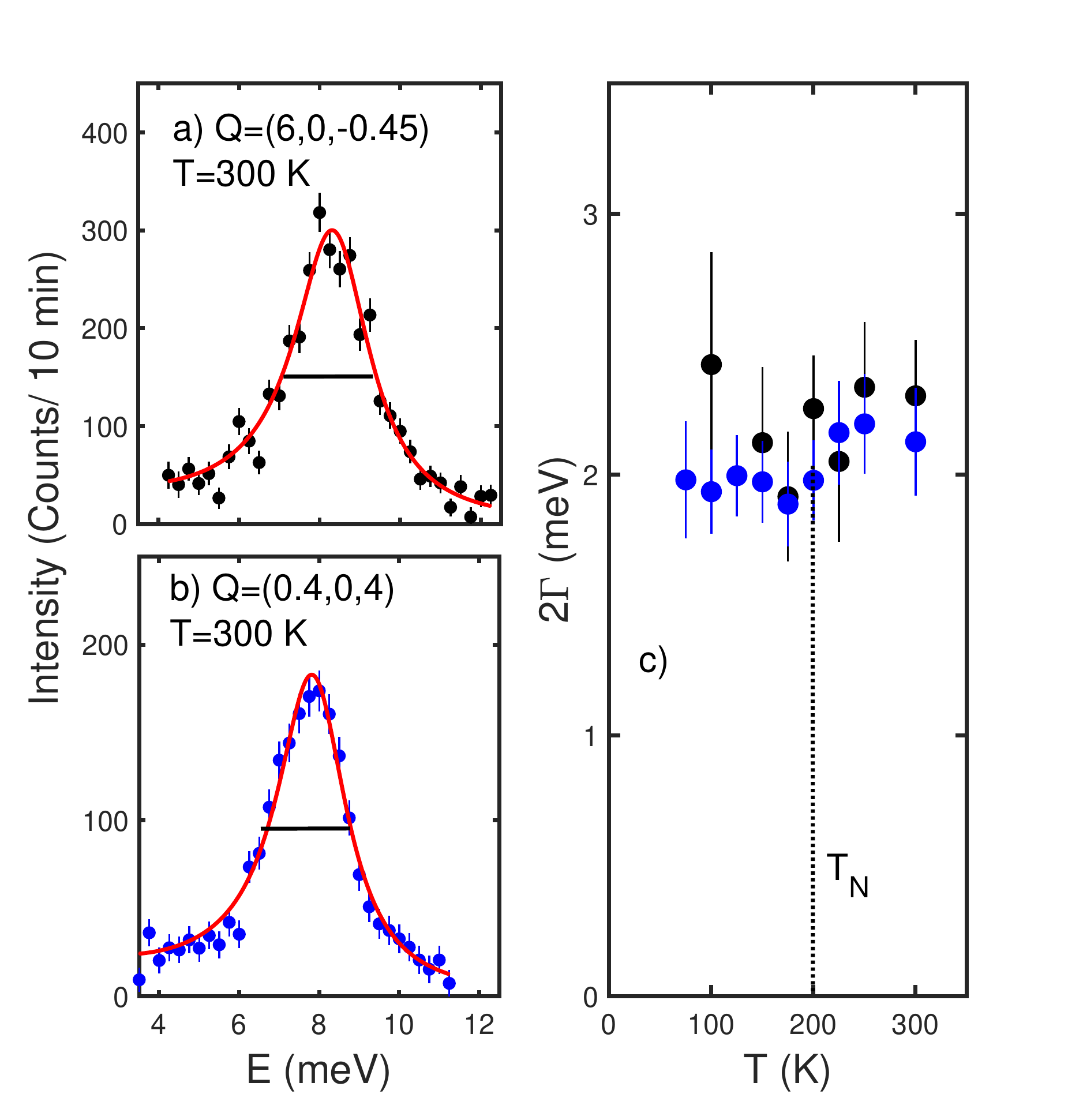}
 \caption{\label{Fig_phonon}  The transverse acoustic phonons propagating along $(a)$ the $c$ and $(b)$ the $a$ axes.  The resolution is depicted by the solid horizontal lines.  $(c)$ illustrates the temperature dependence of both phonon modes.  No measurable anisotropy or temperature dependence is observed.}
\end{figure}
Acoustic phonon measurements were performed on the EIGER triple-axis spectrometer (PSI, Switzerland) \cite{Stuhr}.  The incident neutron beam was monochromated with a vertically focused PG(002) monochromator defining E$_{i}$ and the final energy was fixed to E$_{f}$ = 14.6 meV with a PG(002) analyzer, with the energy transfer defined by E=E$_{i}$-E$_{f}$.  Collimation was set to 80 minutes before and after the sample position and a pyrolytic graphite filter was used after the sample to remove higher order contamination of the neturon beam.   The sample was aligned such that Bragg reflections of the form (H K 0) lay within the horizontal plane.  

Constant momentum cuts through the transverse acoustic phonons propagating along the $c$ and $a$ axes are illustrated in Fig. \ref{Fig_phonon} $(a)$ and $(b)$.  The solid line is a fit to a damped simple harmonic oscillator characterized by the antisymmetric Lorentzian lineshape (Eqn. \ref{eq:flux_dissp}), with $\Omega_{0}$ defining the energy position of the phonon and $\Gamma$ the half-width in energy, inversely proportional to the lifetime $\tau$.  The full width $2\Gamma$ is shown in Fig. \ref{Fig_phonon} $(c)$ for phonons propagating along both directions.   The function defined in Eqn. \ref{eq:flux_dissp} consists of the Bose factor multiplying an odd function which ensures that the scattering cross section satisfies the principle of detailed balance \cite{Shirane}.
 
In analogy to the relaxor ferroelectrics mentioned above where nanoregions of polar order are present, we would expect that phonons travelling along the $c$axis, where antiphase domain boundaries exist, might be damped, and this damping would be temperature-dependent.  Fig. \ref{Fig_phonon} shows three key results; first, the acoustic phonons are not measurably broader than the resolution defined by the spectrometer; second, there is no observable temperature dependence to the linewidth; third, there is no observable anisotropy to the linewidth with phonons travelling along both the $a$ and $c$ axes showing similar responses. While we are constrained by the energy resolution of the spectrometer and also the possibility that any effect from the domains affects lower energy phonons, this result does support the idea presented in this paper that there is no observable structural changes with temperatures that drive the magnetism.  
\section{Fluctuations and Neutron Scattering}
The arguments presented in Sect \ref{APB} rely on knowledge of the exchange parameters in the spin Hamiltonian. In order to validate our model of A phase formation through antiphase boundary freezing we turn to the low energy dynamics.

We present neutron scattering data for CaFe$_{2}$O$_{4}$ showing the magnetic fluctuations in both the high and low temperature phases. Following Refs \cite{Buyers,Sarte,SarteCRO,LaneVI3}, we apply a Green's function formalism to model the low energy excitations in both phases, demonstrating the utility of this method in systems without an orbital degree of freedom. A complete derivation of the Green's function for a general collinear system can be found in Appendix \ref{AppendixTheory}. In applying the Green's function formalism to CaFe$_{2}$O$_{4}$, we show that, in the case of a single-ion Hamiltonian that consists solely of a mean field term, the Green's function collapses to a simple expression allowing the calculation of the dispersion relation and dynamical structure factor. Finally, we fit the neutron scattering data, extracting exchange constants and determining the microscopic spin Hamiltonian.

\begin{figure}[b]
  \includegraphics[width=1\linewidth]{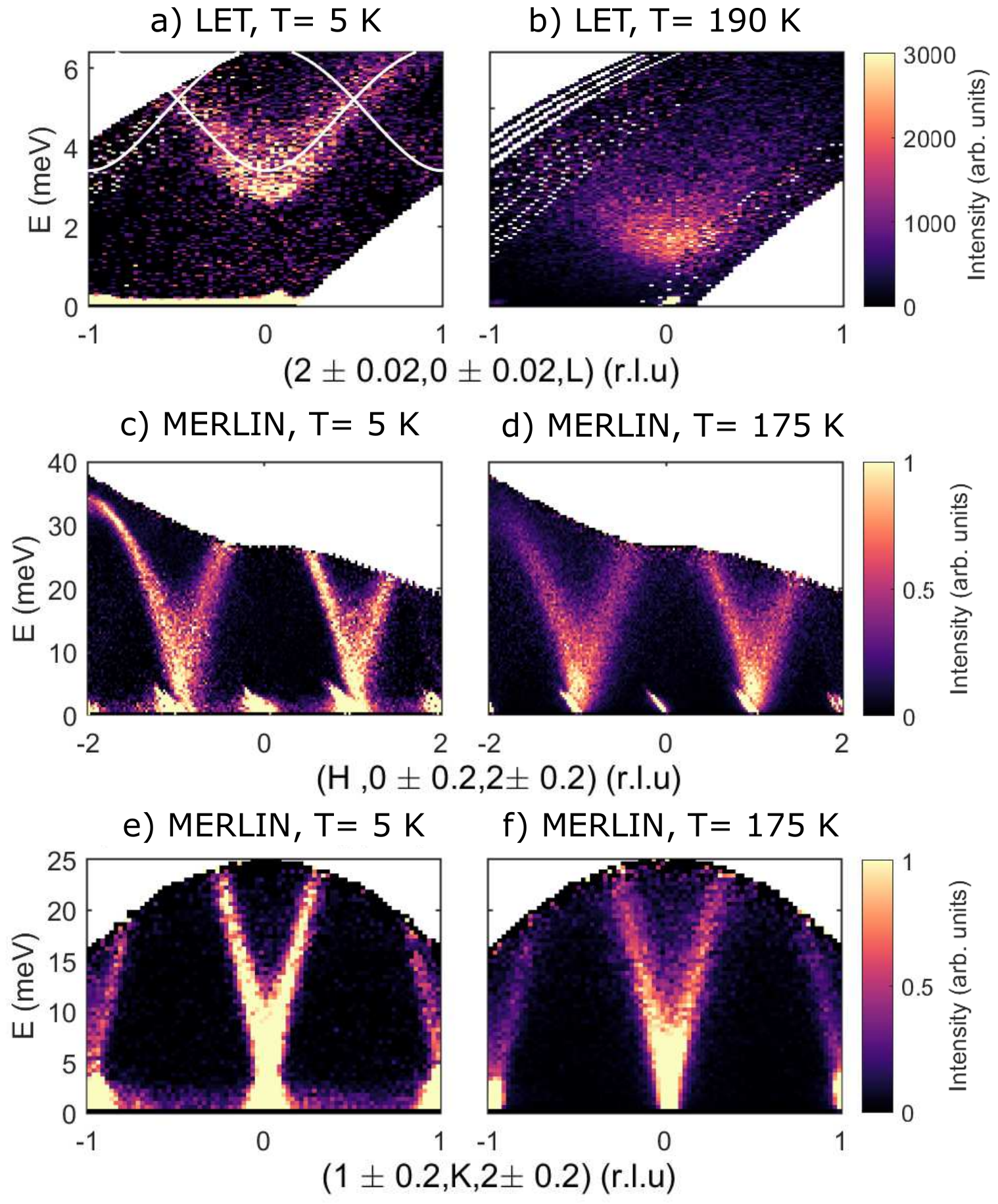}
 \caption{\label{Fig_HKLdisp}  Dispersion along $c^{*}$ at $(a)$ T = 5 K and $(b)$ T = 190 K. Overlaid is the calculated low temperature dispersion, with the fitted exchange constants from Sect. \ref{Spin Hamiltonian}. Spectral weight is concentrated in the mode whose minimum is at $Q=(2,0,0)$. As temperature is increased, the signal broadens and becomes incoherent as correlations along $c$ are lost. $(c)$ Dispersion along H at T = 5 K and $(d)$ 175 K, measured on MERLIN. $(e)$ Dispersion along K at T = 5 K and $(f)$ 175 K. The intensities for both datasets on each instrument have been corrected for the Bose-Einstein population factor.}
\end{figure} 
\subsection{Magnon excitations}

We now discuss the dynamics of CaFe$_{2}$O$_{4}$. Previous studies have shown the presence of rods of diffuse scattering, indicating the presence of antiphase boundaries and revealing the short range nature of correlations along $c$ \cite{Orphan,Solitary}. Despite this, at low temperature, a measurable dispersion along L is observed \cite{Solitary}. We now present data from the cold chopper spectrometer LET \cite{LET} at ISIS Pulsed Neutron and Muon Source (Didcot, UK), concerning the low energy dynamics along L. The incident energy was selected to be $E_{i}= 8$ meV, with the high flux chopper in the 280/140 configuration, giving an elastic resolution of $\Delta E= 0.2$ meV. 

At T = 5 K, a broad gapped low energy mode is observed, extending up to $\sim$ 7 meV as seen in Fig. \ref{Fig_HKLdisp} $(a)$. The gap is $\sim$ 3 meV, in agreement with the data from RITA-II (Fig. \ref{gap} $(b)$). Upon heating to T = 190 K, the gap closes in agreement with the RITA-II data (Fig. \ref{gap} $(c)$), and the scattering broadens becoming incoherent, consistent with the loss of correlations along $c$. It is important to note that our calculations for the dispersion (in both the pure A phase and pure B phase structures) suggest that two modes are present, crossing at $L=0.5$ (Fig. \ref{Fig_HKLdisp} $(a)$). Only one of these modes is observed to carry any spectral weight.
\begin{figure*}
  \includegraphics[width=1\linewidth]{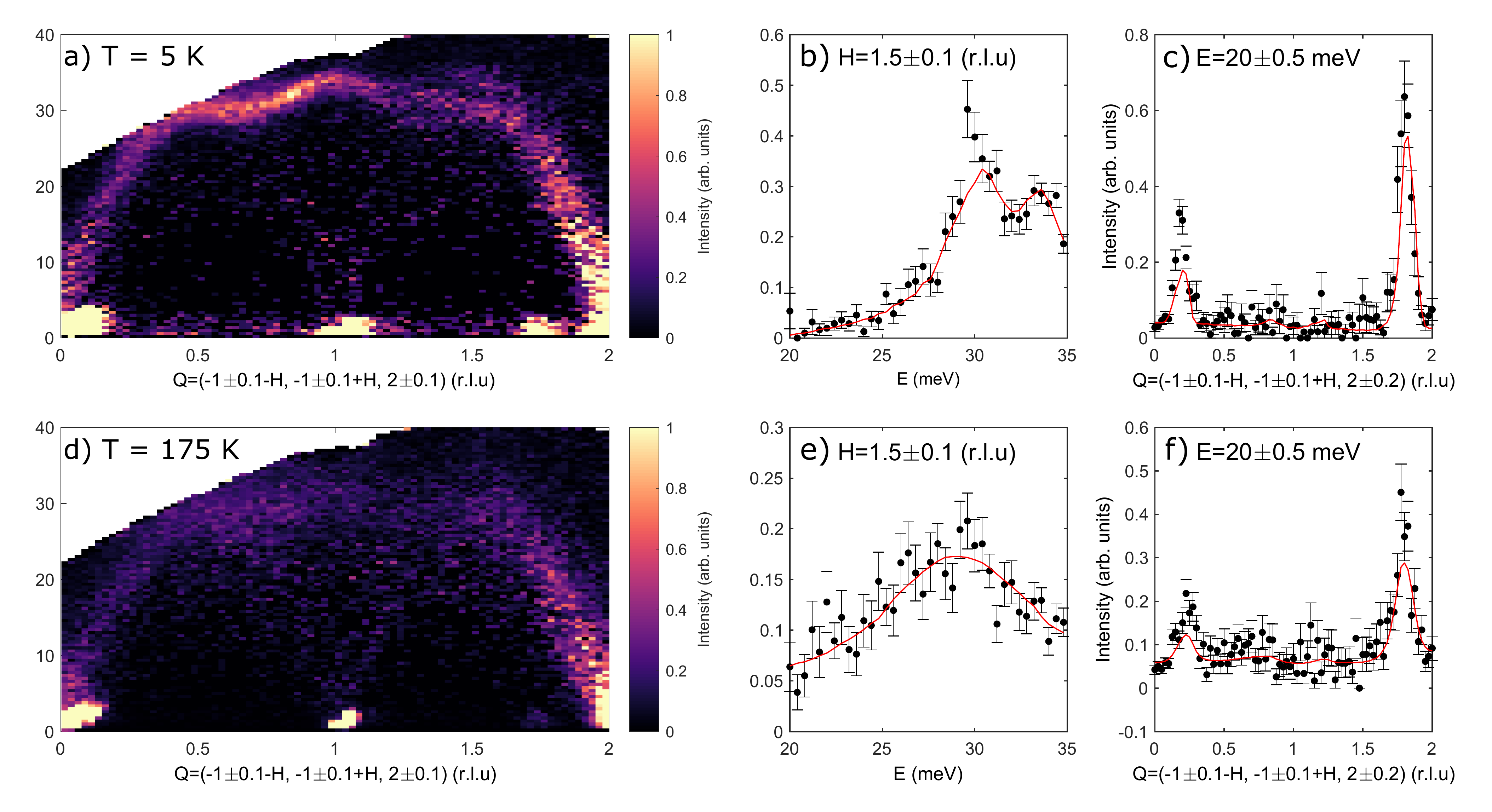}
 \caption{\label{HH2fig}  $(a)$ Cut along [H,-H] at T = 5 K, showing the presence of two modes, with non-trivial structure factor variation. $(b)$ Constant Q cut at the zone boundary. Red line is the fit to the Green's function model. $(c)$ Constant energy cut at E = 27.5 meV with fitted Green's function model. $(d)-(f)$ T = 175 K data showing a broadening of excitations and a small bandwidth renormalization. The intensities for both datasets have been corrected for the Bose-Einstein population factor.}
\end{figure*} 

In order to map out the excitations to higher energies in the (H,K) plane, neutron scattering was carried out on the time-of-flight spectrometer MERLIN at ISIS Neutron and Muon Source (Didcot, UK) \cite{MERLIN}. The sample was cooled to T = 5 K and an incident energy of 70 meV was selected, with a gadolinium chopper spinning at $\nu=300$ Hz, allowing for an elastic resolution of $\Delta E= 3.6$ meV. Strongly dispersive modes were observed along H (Fig. \ref{Fig_HKLdisp} $(c)$), extending up to $\sim$ 35 meV, in agreement with Ref. \cite{Solitary}. Steeply dispersing spin waves were also measured along K (Fig. \ref{Fig_HKLdisp} $(e)$) confirming the three-dimensional nature of the spin waves in this system. 

The sample was then warmed to T = 175 K and the measurement repeated (Fig \ref{Fig_HKLdisp} $(b,d,f)$). The excitations broaden at this temperature, along with a small renormalization of the bandwidth. The dispersion remains qualitatively similar at this temperature, with similar structure factor modulation. Fig. \ref{HH2fig} shows the dispersion along [-H,H]. At low temperature, two modes are seen with two peaks observed at the zone boundary (Fig. \ref{HH2fig} $(b)$). At high temperature, once again the dispersion looks qualitatively similar, but the broadening obscures many of the features seen at low temperature, and the two peaks at the zone boundary are no longer resolved.    We now construct a microscopic spin Hamiltonian to model the spin wave excitations measured at high and low temperatures.

\subsection{Theory}
\label{Spin Hamiltonian}
The Fe$^{3+}$ $(S=\frac{5}{2}, L=0$) ions  in CaFe$_{2}$O$_{4}$ are surrounded by an octahedron of oxygen ions \cite{Decker}. The distorted nature of these octahedra allows for the presence of an easy-axis anisotropy term $\mu \sim B_{2}^{0}$ \cite{AbragamBleaney,Yosida}, breaking spin rotational symmetry and aligning the spins along $b$. The dependence of the anisotropy parameter on the Steven's parameter, $B_{2}^{0}$, suggests an origin for the anomalous temperature dependence of the anisotropy gap presented in Sect. \ref{Tdep}, as the anisotropy parameter is coupled to the local crystalline environment of the Fe$^{3+}$ spins due to the mixing of higher energy multiplets into the ground state orbital singlet \cite{Pryce,Watanabe,Bleaney}. As such, very subtle changes in lattice parameters originating from magnetoelastic coupling, as reported for CaFe$_{2}$O$_{4}$ in Ref. \cite{Songvilay}, can be expected to have a marked effect on the strength of the anisotropy, despite having only a small effect on the Fe-O-Fe bond angle.

The absence of an orbital degree of freedom motivates a spin-only Hamiltonian, 
\begin{equation}
\mathcal{H}=\sum\limits_{ij}\mathcal{J}_{ij}\mathbf{S}_{i}\cdot\mathbf{S}_{j}+\mu \sum\limits_{i}\left(\hat{S}^{z}_{i}\right)^{2}
\end{equation}   
\noindent where $\mu<0$ represents an easy axis single-ion anisotropy. We perform the sum over $j > i$ so that each bond is counted only once. The  existence of two crystallographically inequivalent Fe$^{3+}$ sites, in conjunction with the magnetic order, necessitates the use of an enlarged magnetic supercell of four sites in the B phase. The breaking of inversion symmetry in the low temperature A phase further enlarges the unit cell to eight sites and necessitates the averaging of the spin-inverted structure factors since $S^{+-}(\mathbf{q,\omega})\neq S^{-+}(\mathbf{q,\omega})$. The spin Hamiltonian can be separated into single and inter-ion terms, $\mathcal{H}=\mathcal{H}_{1}+\mathcal{H}_{2}$, by performing a mean field decoupling, $\mathbf{S}_{i,\gamma}\to\langle\mathbf{S}_{i\gamma}\rangle+\delta\mathbf{S}_{i\gamma}$ and discarding terms $\sim \mathcal{O}(\delta\mathbf{S}_{i\gamma})^{2}$  

\begin{subequations}
\label{H1H2}
\begin{align}
  \mathcal{H}_{1}=&\sum_{i\gamma}\hat{S}_{i\gamma}^{z}\left(2\sum_{j\gamma'}J_{ij}^{\gamma\gamma'}\langle \hat{S}^{z}_{j\gamma'}\rangle+2\mu\langle\hat{S}^{z}_{i\gamma}\rangle\right)\\
  \begin{split}
  \mathcal{H}_{2} =& \sum_{ij}^{\gamma\gamma'}\mathcal{J}_{ij}^{\gamma\gamma'}\hat{S}_{i\gamma}^{z}\left(\hat{S}^{z}_{j\gamma'}-2\langle \hat{S}_{j\gamma'}^{z}\rangle\right) \\
    &\qquad+\frac{1}{2}\sum_{ij}^{\gamma\gamma'}\mathcal{J}_{ij}^{\gamma\gamma'}\left(\hat{S}_{i\gamma}^{+}\hat{S}_{j\gamma'}^{-}+\hat{S}_{i\gamma}^{-}S_{j\gamma'}^{+}\right).
  \end{split}
\end{align}
\end{subequations}

\begin{figure}
\includegraphics[width=1\linewidth]{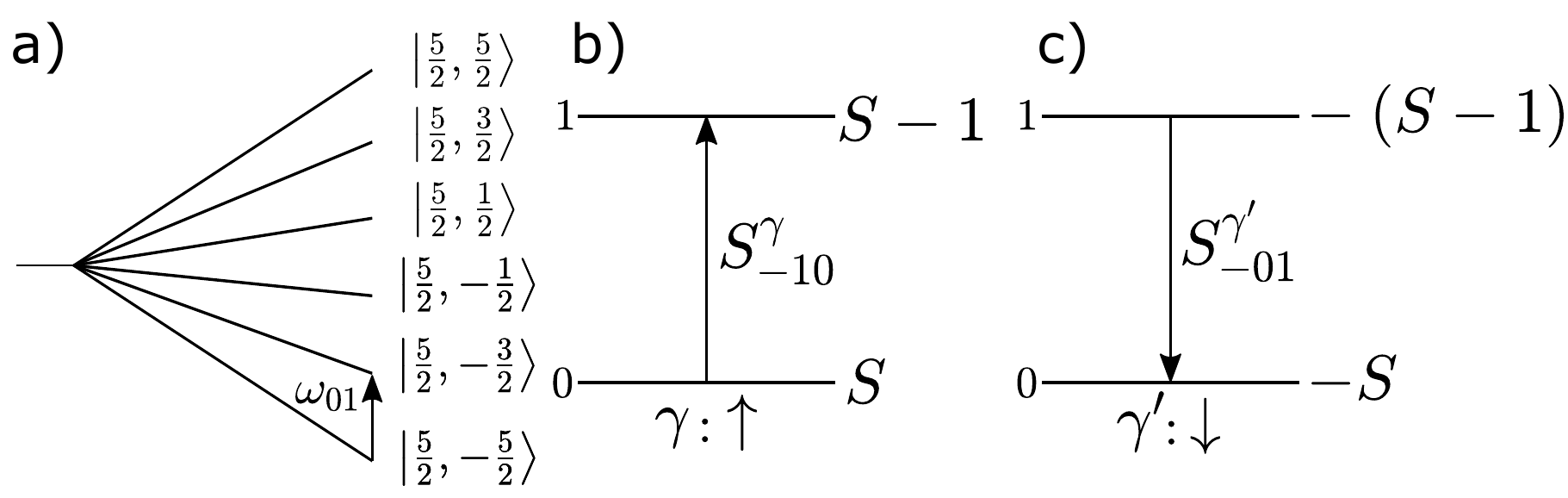}
\caption{\label{Buyers} $(a)$ Visualization of the splitting of the $2S+1$ single-ion energy levels for an $S=5/2, L=0$ ion, due to a molecular mean field. The separation between energy levels is given by $\omega_{01}=\omega_{1}-\omega_{0} $. $(b)$, $(c)$ The effect of the operation of $\hat{S}^{-}$ on sites in the $\uparrow$ $(b)$ and $\downarrow$ $(c)$ state \cite{Buyers_Lecture_Notes}.}
\end{figure}

\noindent The first term is Zeeman term describing the molecular mean field felt by each site $H_{1}=\sum_{i\gamma}h^{MF}_{i\gamma}\hat{S}_{i\gamma}^{z}$. This splits the $2S+1$ degenerate energy levels (Fig. \ref{Buyers}). The commutators $[\hat{S}^{\alpha}_{i'\tilde{\gamma}},\mathcal{H}]$ can be calculated to mean field level; only the transverse elements survive
\begin{subequations}
\begin{align}
[\hat{S}^{+}_{i'\tilde{\gamma}},\mathcal{H}]=\sum_{j\gamma'}A_{i'j}^{\tilde{\gamma}\gamma'}\hat{S}_{j\gamma'}^{+}\\
A_{i'j}^{\tilde{\gamma}\gamma'}=-h^{MF}_{i'\tilde{\gamma}}\delta_{i'j}\delta_{\tilde{\gamma}\gamma'}+2\mathcal{J}_{i'j}^{\tilde{\gamma}\gamma'}\langle \hat{S}_{i'\tilde{\gamma}}^{z} \rangle.
\label{A}
\end{align}
\end{subequations}
\noindent This commutator can be inserted into the Green's function equation of motion (Appendix \ref{AppendixTheory}) to yield the Green's functions, 
\begin{equation}
\begin{split}
\omega G^{+-}_{\tilde{\gamma}\tilde{\gamma}'}(i'j',\omega)=&\langle [\hat{S}^{+}_{i'\tilde{\gamma}},\hat{S}^{-}_{j'\tilde{\gamma}'}]\rangle\\&\qquad+\sum_{j\gamma'}A^{\tilde{\gamma}\gamma'}_{i'j}G_{\gamma'\tilde{\gamma}'}^{+-}(jj',\omega).
\end{split}
\end{equation}
This can be written as
\begin{equation}
\begin{split}
\sum_{j\gamma'}G^{+-}_{\gamma'\tilde{\gamma}'}(jj',\omega)&\left[\omega\delta_{i'j}\delta_{\tilde{\gamma}\gamma'}-A_{i'j}^{\tilde{\gamma}\gamma'}\right]\\&\qquad=\langle [\hat{S}^{+}_{i'\tilde{\gamma}},\hat{S}^{-}_{j'\tilde{\gamma}'}]\rangle.
\end{split}
\label{penultimate}
\end{equation}
\noindent Taking the Fourier transform and performing the summation, we can write Eqn. \ref{penultimate} as a matrix equation. On doing so, the Green's functions take the convenient form
\begin{equation}
\underline{\underline{G}}^{+-}(\mathbf{q},\omega)=\underline{\underline{B}}\left[\mathbb{I}\omega-\underline{\underline{A}}(\mathbf{q})\right]^{-1}.
\label{final}
\end{equation}
\noindent where $B^{\tilde{\gamma}\tilde{\gamma}'}=\delta_{\tilde{\gamma}\tilde{\gamma}'}\langle \hat{S}_{\tilde{\gamma}}\rangle$. This expression for the transverse Green's function is similar to the expression found by dynamical mean field theory \cite{Stripe}, where the correlation function is found from the Landau-Lifshitz equation. The dispersion relation can be found analytically by diagonalizing the matrix $A^{\tilde{\gamma}\tilde{\gamma}'}(\mathbf{q})$ and the Green's function found by calculating the matrix product on the right-hand side of Eqn. \ref{final} on a grid in energy-momentum space. The dynamical structure factor can then be calculated via the fluctuation-dissipation theorem \cite{Boothroyd}
\begin{equation}
S(\mathbf{q},\omega)=-\frac{1}{\pi}\left(1+n(\omega)\right)\textup{Im}G(\mathbf{q},\omega).
\end{equation}

\noindent Expressions for $\underline{\underline{A}}(\mathbf{q})$ and $\underline{\underline{B}}$ can be found in Appendix \ref{Matrices}. We add a small imaginary offset to the energy, $\omega\to \omega +i\epsilon$ to give the intensity peak a finite width. The resultant lineshape takes the form of a Lorentzian of width $2\Gamma=2\epsilon$, which in our low temperature analysis will be set to a value smaller than the instrument resolution.

\begin{figure}
  \includegraphics[width=1\linewidth]{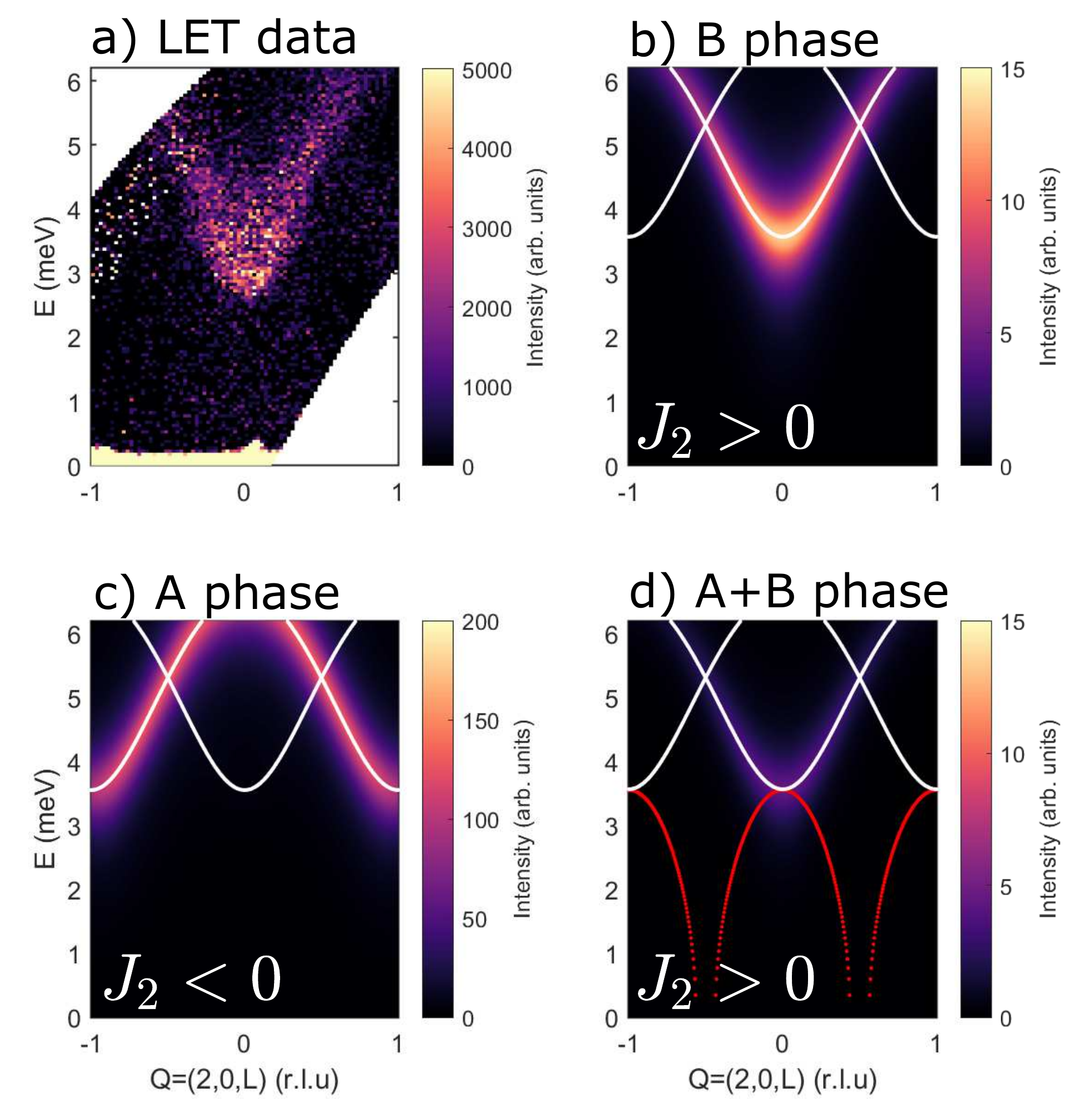}
 \caption{\label{dispalongL} Comparison of calculated dispersion along $c^{*}$ against $(a)$ the data from LET at T = 5 K for a model consisting of $(b)$ B phase order with antiferromagnetic $J_{2}$, $(c)$ A phase order and $(d)$ A and B phase order in a 2:1 ratio. It is clear that for $J_{2}<0$ the mode that is maximal at $Q=(2,0,0)$ lights up, in contrast to the data. Furthermore, for $J_{2}>0$ and two modes are seen and the A phase leads to imaginary eigenvalues (red dispersion) using the fitted values unless $\mu$ is large.     }
\end{figure}

\begin{figure*}
\includegraphics[width=1\linewidth]{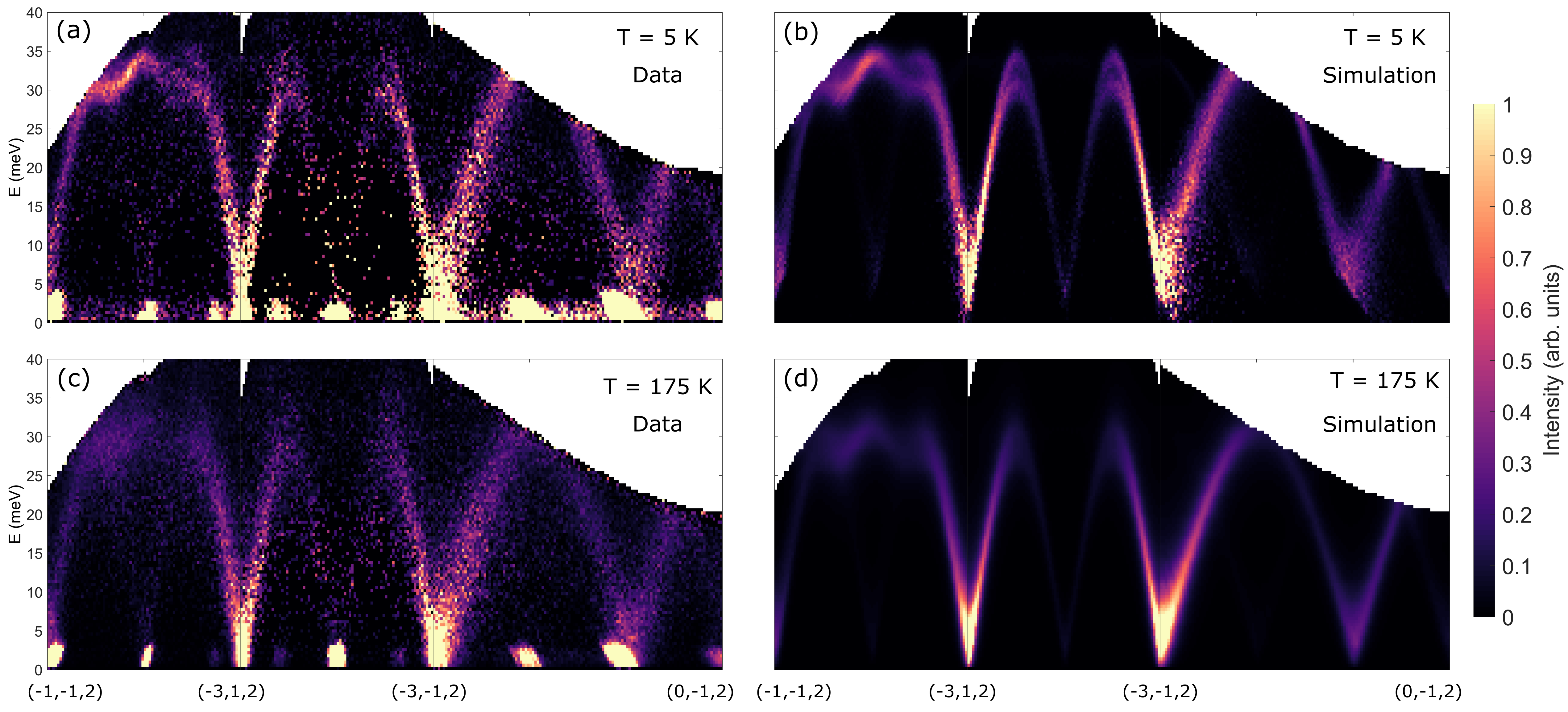}
\caption{\label{Simulation} $(a)$ Constant Q plot of a path through reciprocal space at T = 5 K, showing strongly dispersing excitations in the (H,K) plane. $(b)$ Simulation of the data using the resolution convoluted Green's function model and the fitted parameters. $(c)$ T = 175 K data. $(d)$ Simulation at T = 175 K. The intensities for both datasets have been corrected for the Bose factor. }
\end{figure*}

The six shortest bonds have bond lengths of between 3.01-3.66\AA, therefore it is not clear by distance alone which should be the strongest. The shortest two bonds $J_{1a}$ and $J_{1b}$ are of the same length, both linking parallel spins along $b$, but are crystallographically inequivalent, with $J_{1a}$ and $J_{1b}$ forming the legs of the blue and cyan zig-zag chains respectively. The presence of antiphase domain boundaries, and the near 90$^{\circ}$ Fe-O-Fe exchange path along a shared octahedral edge, indicates that the next two shortest bonds $J_{2a}$ and $J_{2b}$ are likely to be small \cite{Khomskii}. The measurable dispersion along H and L (Fig. \ref{Fig_HKLdisp}) is suggestive of a non-negligible $J_{3}$ and $J_{4}$. Thus in order to write down the minimal physically-motivated model, we must include the six shortest bonds (Fig. \ref{structure fig} $(a)$),  
\begin{equation}
\underline{\underline{\mathcal{J}}}=
\begin{pmatrix}
J_{1a} & J_{2a} & 0 & J_{3} & 0 & J_{4} & 0 &0 \\
J_{2a} & J_{1a} & J_{3} & 0 & J_{4} & 0 & 0 &0\\
0 & J_{3} & J_{1b} & J_{2b} & 0 & 0 & 0 & J_{4} \\
J_{3} & 0 & J_{2b} & J_{1b}& 0 & 0 & J_{4} & 0\\
0 & J_{4} & 0 & 0 & J_{1b} & J_{2b} & 0 & J_{3}\\ 
J_{4} & 0 & 0 & 0 & J_{2b} & J_{1b} & J_{3} & 0 \\ 
0 & 0 & 0 & J_{4} & 0 & J_{3} & J_{1a} &J_{2a} \\
0 & 0 & J_{4} & 0 & J_{3} & 0 & J_{2a} & J_{1a}
\end{pmatrix}.
\label{J}
\end{equation}
\noindent The MERLIN data at T = 5 K were fitted using \texttt{HORACE} \cite{Horace} with values of $\mu$, $J_{2a}$ and $J_{2b}$ fixed. The \texttt{TOBYFIT} package was used to account for the resolution function on MERLIN and contributions from the guide, chopper and moderator were considered. In accordance with our conclusion that the underlying magnetic order is B phase, we used a B phase only model. These parameters were then refined by fitting the LET data using the values obtained from the MERLIN fit. This process was iterated until good agreement was achieved. The effect of taking $J_{2a}\neq J_{2b}$ is to open a gap at the crossing point along L (Fig. \ref{dispalongL}). Such a gap is not seen in the data so we therefore set $J_{2a}=J_{2b}=J_{2}$. The refined values of the exchange constants are listed in Table \ref{exchanges}.
\begin{table}[h]
\caption{\label{exchanges} Fitted exchange constants, $J_{i}$, and anisotropy parameter, $\mu$, for the bonds labelled in Fig. \ref{structure fig} $(a)$.}
\begin{ruledtabular}
\begin{tabular}{ccc}
$J_{i}$ &Distance (\AA) & Value (meV) \\ 
\hline
$J_{1a} $ & 3.018 & 0.03(1)  \\ 
$J_{1b} $ & 3.018 & 0.38(1) \\ 
$J_{2a} $ & 3.077 &0.047(2) \\ 
$J_{2b} $ & 3.096 & 0.047(2) \\
$J_{3}$ & 3.570 & 3.4(3) \\
$J_{4}$ & 3.659 & 3.2(3)\\
$\mu$ & - & -0.035(1)  
\end{tabular}
\end{ruledtabular}
\end{table}
\noindent The dominant exchange couplings were determined to be $J_{3}$ and $J_{4}$, with $J_{2}$ confirmed to be small. The frustrated nature of the bonds $J_{1a}$ and $J_{1b}$ gives rise to the arch-like dispersion at (-2,0,L) (Fig. \ref{HH2fig}), which is well reproduced in our model (Fig. \ref{Simulation}) . Crucially, $J_{2}$ was determined to be small, $J_{2}<0.05$ meV, meaning that the creation of an antiphase boundary carries a small energy cost and thermal fluctuations at high temperature can overcome this barrier, thus explaining the significant fraction of dynamical diffuse scattering \cite{Orphan}.

The effect of nonzero $J_{2}$ warrants some consideration. In one of the two magnetic structures, this bond is expected to be frustrated and hence two copies of the dispersion curve along $c^{*}$ would be expected (with a different gap and bandwidth) if both phases were to contribute to the signal along $c^{*}$. No such duplication of modes is observed (Fig. \ref{dispalongL} $(a)$). Furthermore, we can rule out $J_{2}<0$ since the mode whose minimum lies at $Q=(2,0,0)$ carries the most spectral weight contrary to the simulation (Fig. \ref{dispalongL} $(c)$). This is in agreement with our conclusion based on the neutron diffraction data. Finally, the dispersion along all other directions, along with the measured anisotropy gap put strong constraints on the values of the exchange parameters. With positive $J_{2}$, the magnitude of anisotropy required to stabilize spin waves in the A phase is inconsistent with the observed gap from RITA-II (Fig. \ref{gap}), LET (Fig. \ref{Fig_HKLdisp} $(a,b)$) and MERLIN (Fig. \ref{Fig_HKLdisp} $(c-f)$). We thus conclude that the the low temperature A phase is metastable in our single crystal sample. This phase obtains a long lifetime due to the anisotropy, which prevents the relaxation of the antiphase domain boundary disordered high temperature state into the B phase ground state structure. The extracted exchange constants do not give rise to stable A phase spin waves, as shown in Fig. \ref{dispalongL} $(d)$. Note that in the powder samples, the story need not be identical, with an expected increased value of $\mu$ due to the inevitable finite strain induced by the grinding of the powders, it may in fact be possible to stabilize spin waves in the A phase, despite the lower energy of the B phase configuration. Indeed, the smaller magnitude of the anisotropy gap measured in the low temperature phase \cite{Songvilay} as compared to single crystals (Fig. \ref{gap}) could be due to the suppression of the gap originating from the frustrated $J_{2}$ bond. We now turn our attention to the T = 175 K data.        

We explained earlier in Sect. \ref{structure} how the magnetic A phase arises, not as a distinct phase in the bulk but locally in antiphase boundaries between different B phase domains. This explanation, of the low temperature A phase not existing in bulk, means that its formation is not driven by, for example, a change in sign of $J_{2}$ arising from a change in crystal structure. We note that although the $c$ lattice constant does show some temperature dependence \cite{Songvilay}, this is on the order of $10^{-2}$\AA~and hence is unlikely to affect either the sign or magnitude of $J_{2}$. In the absence of temperature dependence of the exchange parameters, the primary effects of the increased temperature should be the damping and renormalization of the spectrum due to higher order terms in the Dyson series. The damping can be accounted for phenomenologically by increasing the value of $\epsilon$, thereby increasing the Lorentzian linewidth \cite{Chou}. The renormalization takes the form of a reduced spin moment and can be treated straightforwardly by renormalizing the exchange parameters and the anisotropy parameter, $\{ \mathcal{J}_{ij},\mu\}\to\{\gamma\mathcal{J}_{ij},\gamma \mu \}$, where $\gamma$ is some constant between zero and unity. This follows from the fact that $S$ is a dimensionless parameter and so only appears in the dispersion as a multiple of an exchange or anisotropy parameter, allowing us to absorb the renormalization factor into the exchange parameters. As discussed in Sect. \ref{Tdep}, the anisotropy gap shows an anomalous temperature dependence and so we expect a further suppression of $\mu$ beyond that expected by spin moment renormalization alone. Fixing the fitted low temperature exchange parameters and setting $\epsilon=2.5$ meV, we fitted an overall renormalization factor, $\gamma=0.930(4)$, showing excellent agreement with the data (Fig. \ref{Simulation}). At high temperature, the $Q=(1,0,1)$ neutron diffraction peak is not resolution limited and the width is expected to vary in both energy and momentum due to magnon-magnon and magnon-soliton interactions \cite{Allroth}. The value of $\epsilon$ was selected according to the approximate width of the peak at the zone boundary. Using the value of the anisotropy gap at T = 175 K, we diagonalized the Hamiltonian with the fitted renormalized exchange parameters and solved for the high temperature anisotropy parameter $\mu_{175\textrm{K}}=-0.0098(2)$ meV.   

From our fitted exchange parameters we can estimate the Curie-Weiss temperature 
\begin{equation}
\label{CW}
k_{B}\Theta_{CW}=\frac{1}{3}S(S+1)\sum_{n}J_{n},
\end{equation}
\noindent where we sum over nearest neighbors. Due to the inequivalence of $J_{1a}$ and $J_{1b}$, we take the average of their fitted values. The expression above stems from a mean field treatment of the single-ion, and hence we attach a minus sign to the frustrated $J_{1a}$ and $J_{1b}$ bonds. Evaluating Eqn. \ref{CW}, we find $\Theta_{CW} \approx 435$ K. We note that this is significantly larger than that found by Das \textit{et al} \cite{Das}, and much larger than the magnetic ordering temperature of T $\approx$ 200 K. However, one should note that the loss of correlations along $c$, owing to the small value of $J_{2}$ renders CaFe$_{2}$O$_{4}$ quasi-two-dimensional at high temperatures. The absence of spontaneous symmetry breaking for $d\leq 2$ \cite{Mermin} thus makes long range order marginal. Long range order is stabilized by the presence of single-ion anisotropy, however the vanishing of the gap at T = 200 K, due to the cooperative effect of spin moment renormalization and subtle magnetoelastic changes to the local crystalline environment, precludes any long range magnetic order above this temperature. The large Curie-Weiss temperature also explains the relatively modest renormalization of the bandwidth, with an observed moment reduction of $\sim 10$\% at T = 175 K, despite the proximity to the magnetic ordering temperature.       
\section{Conclusion}
In this paper we have shown that the magnetic phase coexistence in CaFe$_{2}$O$_{4}$ can be understood as originating from a parent B phase magnetic order with local A phase order arising due to the freezing of antiphase boundaries which become static below T $\approx$ 100 K. This is consistent with the lack of temperature dependence of the acoustic phonon linewidth, which is sensitive to instabilities in the crystal structure which would lead to changes in the magnetic structure. We have presented neutron scattering data showing the temperature-dependent opening of the anisotropy gap, which stabilizes the low temperature A phase order. We then showed that the magnon excitations are qualitatively consistent at high and low temperature, albeit broadened at high temperature by the dynamical antiphase boundaries. Using a Green's function formalism, we showed that the spectrum can be modeled with the same exchange constants in both phases, save for a renormalization factor at high temperature, but with two different anisotropy parameters owing to the anomalous temperature dependence of $\mu$. The extracted exchange constants are consistent with the picture of antiphase boundary freezing, with a small value of  $J_{2}$. By analysis of the spectrum, it was shown that stable spin waves cannot exist in the A phase and that this phase is metastable, frozen-in at low temperatures due to the growth of the single-ion anisotropy. The anisotropy acts to lock the antiphase boundaries in place, preventing relaxation of the magnetic structure back to the ground state B phase order. 
\begin{acknowledgements}
The authors thank W.J.L. Buyers, P.M. Sarte, M. Songvilay and F. Kr{\"u}ger for useful discussions. Experiments at the ISIS Pulsed Neutron and Muon Source were supported by beamtime allocations RB1710380 and RB1910504 from the Science and Technology Facilities Council. H. L. was co-funded by the ISIS facility development studentship programme. S.W.C. was supported by the DOE under Grant No. DOE: DE-FG02-07ER46382.  E.E.R would like to thank the U.S. Department of Energy, Office of Science (DE-SC0016434) for financial support. This work was supported by the EPSRC, the STFC and the Swiss spallation neutron source (SINQ) (Paul Scherrer Institute, Villigen, Switzerland).
\end{acknowledgements}
\appendix 
\section{Calculation of Green's Functions for a General Collinear System}
\label{AppendixTheory}
In inelastic neutron scattering experiments, the dynamical two-point spin correlation function is probed, which through the fluctuation-dissipation theorem \cite{Boothroyd}, can be related to the system's linear response function. This underlying connection between the the dynamical structure factor, measured in experiment, and the linear response function of the system makes Green's functions the natural language to describe the neutron scattering response. Previous studies \cite{Buyers,Sarte,SarteCRO} have shown the utility of describing systems comprising inequivalent sub-lattices or those exhibiting non-trivial single-ion physics using the Green's function formalism. Here we present a general form of the Green's function formalism for interacting spins within the random phase approximation. The approach presented is similar to the SU($N$) spin wave theory or ``flavor wave" approaches \cite{Muniz,Dong,Elliot,Papanicolaou,Romhanyi,Hasegawa}, however the direct calculation of the Green's function lends itself to the calculation of the neutron response and by way of Wick's theorem, the calculation of magnon-scattering terms.

For a system consisting of interacting spins, the Hamiltonian can be written as
 \begin{equation}
 \mathcal{H}=\sum^{\gamma\gamma'}_{ij}\left\{\mathcal{J}^{\gamma\gamma'}_{ij}\mathbf{S}_{i\gamma}\cdot\mathbf{S}_{j\gamma'}+\mathcal{H}'\left(i,\gamma\right)\right\}, 
 \end{equation}
\noindent where $\mathcal{H}'$ is the single-site Hamiltonian which may include contributions from spin-orbit coupling and crystal-field distortions. Again, we perform the sum over $j>i$ to avoid double counting. The labels $\gamma$ and $\gamma'$ index the sublattice, allowing for the treatment of lattices with multiple atoms within the unit cell. Treating the system at mean-field level, one can separate the Hamiltonian into single-ion, $\mathcal{H}_{1}$ and inter-ion, $\mathcal{H}_{2}$, terms
\begin{subequations}
\begin{align}
\label{spliteqs}
\mathcal{H}=&\mathcal{H}_{1}+\mathcal{H}_{2}\\
\mathcal{H}_{1}=&\sum_{i\gamma}\hat{S}^{z}_{i\gamma} \Big[2\sum_{j\gamma'}\mathcal{J}^{\gamma\gamma'}_{ij}\langle \hat{S}^{z}_{j\gamma'}\rangle\Big]+\sum_{i\gamma}\mathcal{H}'\left(i,\gamma\right) \\
\begin{split}
\mathcal{H}_{2}=&\frac{1}{2}\sum^{\gamma\gamma'}_{ij}\mathcal{J}^{\gamma\gamma'}_{ij}\left(\hat{S}^{+}_{i\gamma}\hat{S}^{-}_{j\gamma'}+\hat{S}^{-}_{i\gamma}\hat{S}^{+}_{j\gamma'}\right)\\ 
&\qquad+\sum^{\gamma\gamma'}_{ij}\mathcal{J}^{\gamma\gamma'}_{ij}\hat{S}^{z}_{i\gamma} \Big[ \hat{S}^{z}_{j\gamma'}-2\langle \hat{S}_{j\gamma'}^{z}\rangle\Big].
\end{split}
\end{align}
\end{subequations}

\noindent The equation of motion for the Green's function, $G^{\alpha\beta}_{\tilde{\gamma}\tilde{\gamma}'}(i'j',t)=-i\Theta(t)\langle [\hat{S}^{\alpha}_{i'\tilde{\gamma}}(t),\hat{S}^{\beta}_{j'\tilde{\gamma}'}]\rangle$, can be written as 
\begin{equation}
\begin{split}
i\partial_{t}G^{\alpha\beta}_{\tilde{\gamma}\tilde{\gamma}'}(i'j',\omega)=&\delta(t)\langle [\hat{S}^{\alpha}_{i'\tilde{\gamma}}(t),\hat{S}^{\beta}_{j'\tilde{\gamma}'}]\rangle\\& \qquad -i\Theta(t)\langle [i\partial_{t}\hat{S}^{\alpha}_{i'\tilde{\gamma}}(t),\hat{S}^{\beta}_{j'\tilde{\gamma}'}]\rangle.
\end{split}
\end{equation}
\noindent Using the Heisenberg equation of motion, the time-dependent spin operator in the second term can be replaced with a commutator, and after a temporal Fourier transform, the Green's function can be recast as a function of energy
\begin{equation}
\begin{split}
\omega G^{\alpha\beta}_{\tilde{\gamma}\tilde{\gamma}'}(i'j',\omega)=&\langle [\hat{S}^{\alpha}_{i'\tilde{\gamma}},\hat{S}^{\beta}_{j'\tilde{\gamma}'}]\rangle+\\& \qquad G_{\tilde{\gamma}\tilde{\gamma}'}([\hat{S}^{\alpha}_{i'\tilde{\gamma}},\mathcal{H}],\hat{S}^{\beta}_{j'\tilde{\gamma}'},\omega).
\end{split}
\label{eqofmot}
\end{equation}

\noindent In the case where $\mathcal{H}_{1}$ consists solely of a mean field term, and $\hat{S}^{z}$ is conserved, the only nonzero Green's functions within the random phase approximation scheme are transverse. In the general case, the commutators, $[\hat{S}^{\alpha}_{i'\tilde{\gamma}},\mathcal{H}]$ must be calculated. We can write the spin operators in terms of the creation operators of the single-ion Hamiltonian
\begin{equation}
\hat{S}^{\alpha}_{i\gamma}=\sum_{pq}S^{\gamma}_{\alpha pq}c^{\dagger}_{p}(i)c_{q}(i)
\end{equation}   

\noindent where $S^{\gamma}_{\alpha pq}=\bra{p}\hat{S}^{\alpha}_{\gamma}\ket{q}$, with the single-ion eigenstates, $\ket{p}$. Using this transformation, the commutator within the Green's function can be calculated. Each term is quartic in bosonic operators but can be decoupled into quadratic terms through the random phase decoupling scheme,
\begin{equation}
\begin{split}
c^{\dagger}_{p}(i)c_{q}(i)c^{\dagger}_{m}(j)c_{n}(j)=&f_{p}(i)\delta_{pq}c^{\dagger}_{m}(j)c_{n}(j)\\& \qquad +f_{m}(j)\delta_{mn}c^{\dagger}_{p}(i)c_{q}(i),
\end{split}
\end{equation}
\noindent with $f_{p}(i')$, the Bose occupation factor of level $p$ on site $i'$. Following the mean-field decoupling, we are left with four terms from the commutator $[\hat{S}^{\alpha}_{i'\tilde{\gamma}},\mathcal{H}]=\sum_{s=1}^{4}\mathcal{C}_{s}$,
\begin{subequations}
\begin{align}
\mathcal{C}_{1}=&\sum^{lkpq}_{j\gamma\gamma'}\phi_{qp}(i')c^{\dagger}_{k}(j)c_{l}(j)S_{\alpha qp}^{\tilde{\gamma}}S_{+pq}^{\gamma}S_{-kl}^{\gamma'}\mathcal{J}_{ij}^{\gamma\gamma'}\\
\mathcal{C}_{2}=&\sum^{lkpq}_{j\gamma\gamma'}\phi_{qp}(i')c^{\dagger}_{k}(j)c_{l}(j)S_{\alpha qp}^{\tilde{\gamma}}S_{-pq}^{\gamma}S_{+kl}^{\gamma'}\mathcal{J}_{ij}^{\gamma\gamma'}\\
\mathcal{C}_{3}=&\sum^{lkpq}_{j\gamma\gamma'}\phi_{qp}(i')c^{\dagger}_{k}(j)c_{l}(j)S_{\alpha qp}^{\tilde{\gamma}}S_{zpq}^{\gamma}S_{zkl}^{\gamma'}\mathcal{J}_{ij}^{\gamma\gamma'}\\
\mathcal{C}_{4}=&\sum_{pq}\left(\omega_{p}-\omega_{q}\right)c^{\dagger}_{q}(i')c_{p}(i')S_{\alpha qp}^{\tilde{\gamma}},
\end{align}
\end{subequations}

\noindent where $\phi_{qp}(i')=(f_{q}(i')-f_{p}(i'))$. Taking advantage of the linearity of the Green's function, we can now insert the commutators into Eqn. \ref{eqofmot}. Setting $\mathcal{J}^{\gamma\gamma'}_{ij}=0$, we recover the single-ion susceptibility

\begin{equation}
g_{\tilde{\gamma}\tilde{\gamma}'}^{\alpha\beta}(\omega)=\sum_{qp}\frac{S^{\tilde{\gamma}}_{\alpha qp}S^{\tilde{\gamma}'}_{\beta pq}\phi_{qp}}{\omega-(\omega_{p}-\omega_{q})},
\end{equation}
\noindent where we assume that single-ion eigenstates are the same for all sites and drop the site index on $f_{p}$. For the calculation of one-magnon processes the sum is performed over transitions to and from the ground state. This step is equivalent to the elimination of the ground state operators in SU($N$) spin wave theory via the local constraint, $b_{0}(i)=b^{\dagger}_{0}(i)=\sqrt{1-\sum^{N-1}_{m=1}b^{\dagger}_{m}(i)b_{m}(i)}$ \cite{Muniz,Hasegawa,Elliot}. After a spatial Fourier transform, the full expression for the Green's function can be found

\begin{equation}
\begin{split}
&G_{\tilde{\gamma}\tilde{\gamma}'}^{\alpha\beta}(\mathbf{q},\omega)=g_{\tilde{\gamma}\tilde{\gamma}'}^{\alpha\beta}(\omega)\\ &\qquad+\sum_{\gamma\gamma'}\mathcal{J}_{\gamma\gamma'}(\mathbf{q})g_{\tilde{\gamma}\gamma}^{\alpha+}(\omega)G_{\gamma'\tilde{\gamma}'}^{-\beta}(\mathbf{q},\omega)\\&\qquad+\sum_{\gamma\gamma'}\mathcal{J}_{\gamma\gamma'}(\mathbf{q})g_{\tilde{\gamma}\gamma}^{\alpha-}(\omega)G_{\gamma'\tilde{\gamma}'}^{+\beta}(\mathbf{q},\omega)  \\&\qquad+2\sum_{\gamma\gamma'}\mathcal{J}_{\gamma\gamma'}(\mathbf{q})g_{\tilde{\gamma}\gamma}^{\alpha z}(\omega)G_{\gamma'\tilde{\gamma}'}^{z\beta}(\mathbf{q},\omega).
\end{split}
\label{fulleq}
\end{equation}
\noindent The symmetry of the single-ion environment can significantly simplify Eqn. \ref{fulleq}. For octahedral and tetragonal crystalline environments $S_{+pq}=S_{-pq}=0$, even in the presence of a trigonal or tetragonal distortion. However, a rhombic distortion, for example, gives rise to nonzero terms in these matrices \cite{Buyers}. Assuming  a sufficiently symmetric single-ion environment, the $g_{\gamma\gamma'}^{++}(\omega)$ and $g_{\gamma\gamma'}^{--}(\omega)$ terms vanish and the three nonvanishing Green's functions can be written as

\begin{subequations}\label{G+-}
    \begin{align}
        G_{\tilde{\gamma}\tilde{\gamma}'}^{+-}(\mathbf{q},\omega) = & g_{\tilde{\gamma}\tilde{\gamma}'}^{+-}(\omega) \nonumber \\
       &   +\sum_{\gamma\gamma'}\mathcal{J}_{\gamma\gamma'}(\mathbf{q})g_{\tilde{\gamma}\gamma}^{+-}(\omega)G_{\gamma'\tilde{\gamma}'}^{+-}(\mathbf{q},\omega) \\
        G_{\tilde{\gamma}\tilde{\gamma}'}^{-+}(\mathbf{q},\omega) = & g_{\tilde{\gamma}\tilde{\gamma}'}^{-+}(\omega) \nonumber \\
       &   +\sum_{\gamma\gamma'}\mathcal{J}_{\gamma\gamma'}(\mathbf{q})g_{\tilde{\gamma}\gamma}^{-+}(\omega)G_{\gamma'\tilde{\gamma}'}^{-+}(\mathbf{q},\omega) \\
               G_{\tilde{\gamma}\tilde{\gamma}'}^{zz}(\mathbf{q},\omega) = & g_{\tilde{\gamma}\tilde{\gamma}'}^{zz}(\omega) \nonumber \\
       &   +2\sum_{\gamma\gamma'}\mathcal{J}_{\gamma\gamma'}(\mathbf{q})g_{\tilde{\gamma}\gamma}^{zz}(\omega)G_{\gamma'\tilde{\gamma}'}^{zz}(\mathbf{q},\omega). 
    \end{align}
\end{subequations}

\begin{figure}
\includegraphics[width=1\linewidth]{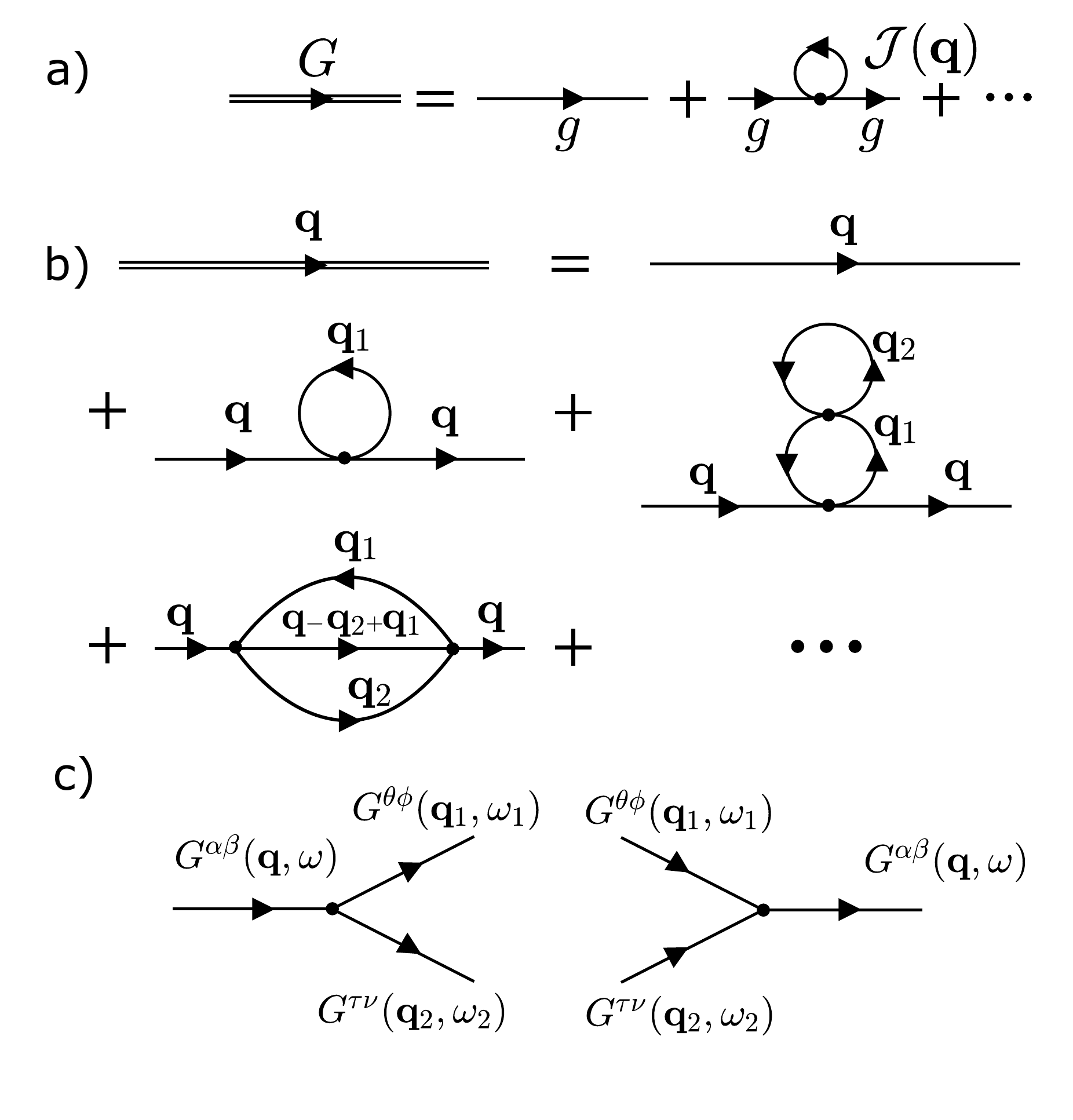}
\caption{\label{Feynman} $(a)$ Feynman diagrams showing the Dyson series structure of the expression for the Green's function obtained in Eqn. \ref{fulleq}. $(b)$ Dyson series for the magnon Green's function showing the first and second order perturbative corrections to the magnon propagator for a collinear spin system. $(c)$ Decay and source channels for three magnon interactions. }
\end{figure}

\noindent In RPA, fluctuations on different sites are taken to be uncorrelated so that $g^{+-}_{\gamma\gamma'}=0$, for $\gamma\neq \gamma'$. These coupled equations can be solved analytically and summed in order to calculate the total Green's function for the system \cite{Sarte}. The coupled equations are most straightforwardly solved as matrix equations,
\begin{subequations}
\begin{align}
\underline{\underline{G}}^{+-}= &\underline{\underline{g}}^{+-}+\underline{\underline{g}}^{+-}\cdot\underline{\underline{\mathcal{J}}}\cdot\underline{\underline{G}}^{+-}\\
\underline{\underline{G}}^{-+}= &\underline{\underline{g}}^{-+}+\underline{\underline{g}}^{-+}\cdot\underline{\underline{\mathcal{J}}}\cdot\underline{\underline{G}}^{-+}\\
\underline{\underline{G}}^{zz}= &\underline{\underline{g}}^{zz}+2\underline{\underline{g}}^{zz}\cdot\underline{\underline{\mathcal{J}}}\cdot\underline{\underline{G}}^{zz}.
\end{align} 
\end{subequations} 
\noindent By examining Eqn. \ref{G+-}, the reasoning behind our decision to label $g^{\alpha\beta}(\omega)$ as the single-ion susceptibility becomes clear. The equations for $\underline{\underline{G}}$ have the form of a Dyson equation \cite{Dyson}, where the single-ion susceptibility plays the role of the bare propagator, and the self-energy is the Fourier transform of the exchange interaction, $\mathcal{J}(\mathbf{q})$ (Fig. \ref{Feynman} $(a)$). The single ion Hamiltonian was treated according to the harmonic approximation and the inter-ion interaction can be considered as a first order perturbative correction, which we decoupled in the direct channel by way of the mean field decoupling, from which the random phase approximation derives its name.

The magnon propagator itself satisfies a Dyson equation and hence by performing this calculation in terms of Green's functions, we can go beyond the single magnon picture and calculate the effect of magnon-magnon scattering on the inelastic neutron response by using Feynman diagram rules. The effect of these higher order terms is, at one-loop level, to dress the magnon propagator with a self energy depending on the magnon density. At higher orders in perturbation theory we add corrections this self energy. Two irreducible topologically distinct Feynman diagrams can be written down \cite{Rastelli}, with their interaction potentials calculable from the Dyson Maleev or Holstein Primakoff Hamiltonian \cite{Harris, Bayrakci}, since the single-ion physics has been treated in the bare magnon propagator, and the spin correlator can equally be written in terms of magnon creation operators. These higher order terms each carry a factor of $1/S$ per vertex \cite{Harris} and are small for $S\to\infty$. The two-loop diagram (Fig. \ref{Feynman} $(b)$) provides a real contribution to the self energy \cite{Rastelli} and hence it renormalizes the spectrum. The next diagram contains a real part and an imaginary part which represents a damping term. This gives an energy broadening to the magnon linewidth, which depends on momentum \cite{Rastelli}. 

In some cases, there are further vertices that should be considered. In systems where $g^{zz}(\omega)$ is nonzero, that is to say, $\langle \hat{S}^{z} \rangle$ is not conserved, the extension beyond the harmonic approximation yields vertices where three lines meet. These terms are absent from spin rotationally symmetric models, as is evident from the absence of cubic terms in the Holstein Primakoff Hamiltonian for collinear systems. Similar terms appear in noncollinear systems where SO(2) symmetry is broken \cite{Zhitomirsky}. These terms represent two magnon decays into a single magnon or vice versa. Using Wick's theorem, any $n$-point correlator can be decomposed into a sum of all possible contractions of the two-point correlators, thus we can calculate the effect of these decay vertices from the Green's functions evaluated by the approach outlined here (Eqn. \ref{fulleq}),
\begin{widetext}
\begin{equation}
\begin{split}
S(\mathbf{q},\omega)\propto -\textrm{Im}\Bigg[ \sum_{\theta\phi\tau\nu}\Big\{\int \mathrm{d}\mathbf{q}_{1}\int \mathrm{d}\mathbf{q}_{2}\int \mathrm{d}\omega_{1}\int \mathrm{d}\omega_{2}G^{\theta\phi}(\mathbf{q}_{1},\omega_{1})G^{\tau\nu}(\mathbf{q}_{2},\omega_{2})\\
\times\delta(\omega-\omega_{1}-\omega_{2})\delta(\mathbf{q}-\mathbf{q}_{1}+\mathbf{q}_{2})\Big\}\Bigg],
\end{split}
\end{equation}
\end{widetext}
\noindent where we have ensured conservation of momentum and energy. The decay amplitudes are governed by kinematics. In particular, it has been argued that the the longitudinal mode, $G^{zz}$ is particularly susceptible to decay into two lower energy transverse spin waves \cite{Sachdev}. Though the calculation of these terms has been simplified by formulating the spin wave calculation in terms of Green's functions, it still remains a formidable task to evaluate this integral in systems which disperse in more than one direction. This is especially true for the fitting of neutron scattering data measured on a time-of-flight spectrometer where one typically integrates over a finite window in momentum space to improve statistics.
\section{Expressions for $A^{\tilde{\gamma}\tilde{\gamma}'}$ and $B^{\tilde{\gamma}\tilde{\gamma}'}$}
\label{Matrices}
The calculation of the transverse Green's function (Eqn. \ref{final}) requires knowledge of the matrix elements, $A^{\tilde{\gamma}\tilde{\gamma}'}$ and $B^{\tilde{\gamma}\tilde{\gamma}'}$. The low temperature A phase structure necessitates an eight site model, owing to the broken inversion symmetry. The matrix $\underline{\underline{B}}$ encodes the magnetic structure of the ground state in each phase (Fig. \ref{structure fig} $(a,b)$), and can be written as
\begin{subequations}
\begin{align}
\underline{\underline{B}}_{A}=
\begin{pmatrix}
-S&0&0&0&0&0&0&0\\
0&-S&0&0&0&0&0&0\\
0&0&S&0&0&0&0&0\\
0&0&0&S&0&0&0&0\\
0&0&0&0&S&0&0&0\\
0&0&0&0&0&S&0&0\\
0&0&0&0&0&0&-S&0\\
0&0&0&0&0&0&0&-S\\
\end{pmatrix}
\\
\underline{\underline{B}}_{B}=
\begin{pmatrix}
S&0&0&0&0&0&0&0\\
0&-S&0&0&0&0&0&0\\
0&0&S&0&0&0&0&0\\
0&0&0&-S&0&0&0&0\\
0&0&0&0&S&0&0&0\\
0&0&0&0&0&-S&0&0\\
0&0&0&0&0&0&S&0\\
0&0&0&0&0&0&0&-S\\
\end{pmatrix}
\end{align}
\end{subequations}
\noindent in the A and B phase respectively, where $S=\frac{5}{2}$. Note that the site labeling has been chosen to match that of Eqn. \ref{J}. The matrix for the B phase can be written as a scalar matrix whose elements are 4 $\times$ 4 matrices, reflecting the inversion symmetry of the B phase magnetic structure. The mean molecular field can be calculated by expanding the spin operators around their expectation values (Eqn. \ref{H1H2}). The presence of the bond inequivalence, $J_{1a}\neq J_{1b}$, and $J_{2a}\neq J_{2b}$, gives rise to two molecular mean field terms, 

\begin{subequations}
\begin{align}
h_{a}^{MF}=&-2J_{1a}S+2J_{2a}S+2J_{3}S+2J_{4}S-2\mu S\\
h_{b}^{MF}=&-2J_{1b}S+2J_{2b}S+2J_{3}S+2J_{4}S-2\mu S
\end{align} 
\end{subequations} 

\noindent where the minus sign in front of the first term reflects the fact that $J_{1a}$ and $J_{1b}$ couple parallel spins, whilst the other exchanges couple spins that are anti-parallel. The matrix $A^{\tilde{\gamma}\gamma'}(\mathbf{q})$ consists of a contribution from the molecular mean field and from the Fourier transform of the exchange interaction (Eqn. \ref{J}), $\underline{\underline{A}}=\underline{\underline{A}}^{MF}+\underline{\underline{A}}^{exch}$. Its matrix elements can be calculated using Eqn. \ref{A}
\begin{widetext}
\begin{subequations}
\begin{align}
\underline{\underline{A}}^{MF}_{A}=&
\begin{pmatrix}
-h^{MF}_{a}&0&0&0&0&0&0&0\\
0&-h^{MF}_{a}&0&0&0&0&0&0\\
0&0&h^{MF}_{b}&0&0&0&0&0\\
0&0&0&h^{MF}_{b}&0&0&0&0\\
0&0&0&0&h^{MF}_{b}&0&0&0\\
0&0&0&0&0&h^{MF}_{b}&0&0\\
0&0&0&0&0&0&-h^{MF}_{a}&0\\
0&0&0&0&0&0&0&-h^{MF}_{a}
\end{pmatrix}\\
\underline{\underline{A}}^{MF}_{B}=&
\begin{pmatrix}
h^{MF}_{a}&0&0&0&0&0&0&0\\
0&-h^{MF}_{a}&0&0&0&0&0&0\\
0&0&h^{MF}_{b}&0&0&0&0&0\\
0&0&0&-h^{MF}_{b}&0&0&0&0\\
0&0&0&0&h^{MF}_{b}&0&0&0\\
0&0&0&0&0&-h^{MF}_{b}&0&0\\
0&0&0&0&0&0&h^{MF}_{a}&0\\
0&0&0&0&0&0&0&-h^{MF}_{a}
\end{pmatrix}.
\end{align} 
\end{subequations}
\end{widetext}
\noindent Finally, the contribution from the exchange term can be calculated by taking the product of  $2\underline{\underline{B}}$ and the Fourier transform of Eqn. \ref{J}, which is the same for both phases,
\begin{equation}
\underline{\underline{A}}_{A/B}^{exch}=2\underline{\underline{B}}_{A/B}\cdot \underline{\underline{\mathcal{J}}}(\mathbf{q}).
\end{equation} 
\noindent By diagonalizing the matrix $\underline{\underline{A}}$ one can obtain an expression for the spin wave dispersion and the Green's function can be calculated using Eqn. \ref{final}. 
\bibliography{CFO_RPA_bib}
\end{document}